\documentclass[a4paper,12pt]{article}
\usepackage{graphicx} 
\usepackage{xcolor}
\usepackage{url}
\usepackage{caption}
\usepackage{subcaption}
\usepackage{array}
\usepackage{tabularx}
\usepackage{comment}
\usepackage[numbers]{natbib}
\usepackage{booktabs}
\usepackage{placeins}
\usepackage{multirow}

\usepackage{geometry}
\title{On the energy efficiency of sparse matrix computations on multi-GPU clusters\thanks{This work was partially supported by: Spoke 6 ``Multiscale Modelling \& Engineering Applications” of the Italian Research
 Center on High-Performance Computing, Big Data and Quantum Computing (ICSC) funded by
 MUR-NextGenerationEU (NGEU); the ``Energy Oriented Center
 of Excellence (EoCoE III): Fostering the European Energy Transition with Exascale” EuroHPC
 Project N. 101144014, funded by European Commission (EC).}}
 \author{Massimo Bernaschi\thanks{Institute for Applied Computing ``Mauro Picone'', National Research Council of Italy, Via dei Taurini, 19, Rome 00185, Italy.
  (massimo.bernaschi@cnr.it, alessandro.celestini@cnr.it, giorgio.richelli@gmail.com)} \and Alessandro Celestini$^\dag$ \and Giorgio Richelli$^\dag$
\and Pasqua D'Ambra\thanks{Institute for Applied Computing ``Mauro Picone'', National Research Council of Italy, Via Pietro Castellino, 111, Naples 80131, Italy.
  (pasqua.dambra@cnr.it)} }

\begin{document}

\maketitle

\begin{abstract}
We investigate the energy efficiency of a library designed for parallel computations with sparse matrices. The library leverages high-performance, energy-efficient Graphics Processing Unit (GPU) accelerators to enable large-scale scientific applications. Our primary development objective was to maximize parallel performance and scalability in solving sparse linear systems whose dimensions far exceed the memory capacity of a single node.
To this end, we devised methods that expose a high degree of parallelism while optimizing algorithmic implementations for efficient multi-GPU usage. Previous work has already demonstrated the library’s performance efficiency on large-scale systems comprising thousands of NVIDIA GPUs, achieving improvements over state-of-the-art solutions.
In this paper, we extend those results by providing energy profiles that address the growing sustainability requirements of modern HPC platforms. We present our methodology and tools for accurate runtime energy measurements of the library’s core components and discuss the findings. Our results confirm that optimizing GPU computations and minimizing data movement across memory and computing nodes reduces both time-to-solution and energy consumption. Moreover, we show that the library delivers substantial advantages over comparable software frameworks on standard benchmarks.
\end{abstract}

\section{Introduction}
\label{intro}

Sustainability is becoming a major concern in large-scale scientific computing. While high-performance computing platforms advance in computational and communication capabilities, power constraints pose significant challenges, impacting operational costs and system reliability up to the point of being a pivotal issue for organizations and nations alike. The energy consumption of data centers received considerable attention in the International Energy Agency’s 2025 report. In particular, the rapid expansion of data centers in countries such as China and the United States\footnote{In the United States, data centers are expected to account for approximately $11.7\%$ of the country's total electricity demand by 2030.} makes the sector one of the major drivers of electricity demand growth, with substantial implications for national energy landscapes~\cite{energy2025}.
Achieving optimal efficiency in power usage is essential for enabling sustainable HPC infrastructures by minimizing operational costs~\cite{suarez2025}.
The rise of Green IT, which advocates for environmentally sustainable computing, has intensified efforts to improve the energy efficiency of HPC systems. The Green500 list~\cite{green500}, which ranks the world's most energy-efficient supercomputers based on their performance in FLOPS per Watt, underscores this growing emphasis on energy-efficient supercomputing, highlighting systems with superior performance-to-power ratios. 

Notably, the adoption of Graphics Processing Units (GPUs) boosted energy-efficient computing, particularly in applications that benefit from parallel processing~\cite{bridges2016understanding,Anztetal2017,suarez2025}. Compared to Central Processing Units (CPUs), GPUs offer superior computational and energy efficiency, especially for high-throughput, high-latency workloads such as scientific machine learning and image processing. As a result, most modern HPC platforms are hybrid systems made of CPUs and GPUs tightly coupled to each other.
The widespread adoption of GPU-accelerated HPC systems is well reflected in the Top500~\cite{top500} and Green500 rankings of the world’s fastest and most energy-efficient supercomputers. Specifically, in the most recent Green500 list, all of the top ten supercomputers are equipped with GPUs, eight of which feature NVIDIA chips.
So, it is essential to assess the energy consumption of GPU sub-systems in today’s power-constrained computing landscape.
To address this challenge, hardware and software strategies have been developed to optimize power usage in supercomputing applications~\cite{Agostaetal2022,filgueras2024textarossa}. 

The primary goal of power management in high-performance computing is to minimize energy consumption while maintaining computational performance within predefined limits. Higher energy efficiency can be achieved by reducing either average power consumption or execution times. 
Research indicates that source-code transformations and application-specific optimizations can significantly enhance GPU resource utilization, performance, and energy efficiency~\cite{Anztetal2017}. Substantial energy savings can be achieved by refining GPU implementations and addressing performance bottlenecks. The authors of~\cite{suarez2025} highlight that even greater efficiency gains are possible by utilizing optimized numerical libraries developed by experts to leverage hardware features and reduce runtimes.

In this paper, we address the critical issue of energy consumption in the core functionalities of \texttt{BootCMatchGX}, a parallel library for sparse matrix computations. 
The library is the outcome of a numerical software development project focused on designing and implementing scalable sparse linear solvers and preconditioners for NVIDIA GPU-accelerated supercomputers. It includes comprehensive support for Krylov subspace methods, incorporating Sparse Basic Linear Algebra operations—such as the sparse matrix-vector product (SpMV)—that are specifically optimized for efficiency on heterogeneous clusters. Additionally, it provides essential operations for Algebraic MultiGrid (AMG) preconditioners.
We present a methodology for analyzing the energy consumption profiles of both the fundamental operations and the overall linear solver, utilizing software tools that enable access to internal hardware sensors~\cite{bridges2016understanding}.
While previous studies have mainly addressed its performance and scalability, here we complement those results with a detailed energy analysis. Our aim is to provide a comprehensive assessment of both runtime efficiency and energy consumption, which is crucial for sustainable high-performance computing.
The main contributions of this paper can be summarized as follows:
\begin{itemize}
\item We present a methodology for fine-grained power measurement based on internal CPU and GPU sensors, integrating the LIKWID toolset with our powerMonitor utility.

\item We provide detailed energy profiles of the library’s fundamental building blocks, including sparse matrix–vector multiplication (SpMV), Conjugate Gradient (CG), and Preconditioned Conjugate Gradient (PCG) solvers, offering a holistic view of performance-to-energy trade-offs.

\item We carry out extensive strong and weak scalability experiments using up to 64 NVIDIA GPUs, and compare  \texttt{BootCMatchGX} against state-of-the-art frameworks such as Ginkgo and NVIDIA AmgX.
\end{itemize}

Our results show that \texttt{BootCMatchGX} consistently achieves lower execution times and reduced dynamic energy consumption compared to Ginkgo. In addition, in the PCG case, \texttt{BootCMatchGX}  outperforms NVIDIA AmgX due to the improved convergence properties of its preconditioner and fine-tuning GPU implementation of basic operations. These findings confirm the effectiveness of algorithmic optimizations and communication-reduction strategies for both performance and energy sustainability.

The remainder of this paper is organized as follows. Section \ref{relwork} collocates this work in the current literature on the energy efficiency aspects of sparse matrix computations on GPU-accelerated systems. Section \ref{bootcmatchgx} presents the design and main features of the BootCMatchGX library. Section \ref{energy} describes the methodology and tools employed for power and energy measurements. Section \ref{results} reports and compares the experimental results for SpMV, Conjugate Gradient, and Preconditioned Conjugate Gradient computations on multi-GPU clusters. Finally, Section \ref{concl} summarizes the main findings and outlines directions for future work.

\section{Energy efficiency in sparse matrix computations for GPU-accelerated systems}
\label{relwork}

Sparse matrix computations are the core of many scientific applications, ranging from traditional simulation models based on Partial Differential Equations (PDEs) to more recent scientific machine learning approaches~\cite{Alya2024,Keyes2024}. In particular, iterative Krylov solvers are the methods of choice for solving large, sparse linear systems involving hundreds of billions of equations, problems that can be efficiently handled using current petascale and pre-exascale high-performance computing (HPC) systems. Profiling the energy consumption of such computations, alongside traditional performance metrics, provides a more comprehensive understanding of the overall efficiency of existing software frameworks.

In general, sparse matrix computations exhibit a memory-bound behavior due to their low operational intensity. Consequently, they are more sensitive to memory and network bandwidth limitations than to the floating-point throughput of the underlying architecture. This characteristic presents a significant challenge on heterogeneous systems with deep memory hierarchies, where the energy cost of data movement is often orders of magnitude higher than that of performing a double-precision floating-point operation~\cite{Kestoretal2013,Dongarra2022,Delestracetal2024}.
This has led to growing interest in analyzing the energy efficiency of sparse matrix computations on GPU-accelerated systems. 

Several studies have conducted experiments and comparisons across different implementations of key computational kernels—such as sparse matrix-vector multiplication (SpMV)—on a range of test cases. In~\cite{Anztetal2017}, the authors compare the performance and energy efficiency of SpMV implementations from cuSPARSE and MAGMA for GPUs, as well as Intel’s MKL for multicore CPUs, on the Swiss supercomputer Piz Daint. Their findings show that optimized GPU code can significantly enhance both the computational and energy efficiency of scientific applications. A comprehensive review of state-of-the-art SpMV implementations on GPUs, including the use of machine learning techniques to select the most efficient method in terms of runtime and energy consumption, is provided in~\cite{Dufrechou2021}. 

In~\cite{Anztetal2012} authors analyzed the power consumption of GPU-accelerated Generalized Minimal Residual (GMRES) solvers enhanced with preconditioning, mixed-precision iterative refinement, and CPU-focused power-saving techniques such as idle-wait periods and Dynamic Voltage and Frequency Scaling (DVFS). While these methods can reduce energy consumption by $6-10\%$, they are less effective for solvers like CG, where the SpMV dominates and leaves little idle time for the CPU.
A more recent study~\cite{Anztetal2024} introduced batched sparse and mixed-precision linear algebra interfaces tailored for applications involving many small-scale systems. By grouping operations into batches and exploiting mixed-precision arithmetic, the approach improves GPU utilization and achieves notable reductions in energy-to-solution.
Furthermore,~\cite{Thomasetal2024} presents a detailed analysis of the performance and energy footprint of the CG method combined with Gauss-Seidel-based preconditioners, specifically designed for GPUs, across various GPU architectures.

\section{\texttt{BootCMatchGX} library}
\label{bootcmatchgx}

\texttt{BootCMatchGX} is the latest development in a mathematical software project aimed at designing new methods and efficient implementations of iterative Krylov solvers and algebraic multigrid preconditioners for solving sparse linear systems. It expands and enhances the sequential library \texttt{BootCMatch}~\cite{BCMTOMS2018} and its Nvidia GPU version, \texttt{BootCMatchG}~\cite{BCMParco2020,BCMswimpact2020}. Its design is driven by the need to scale the library to thousands of GPUs, enabling the solution of systems with many billions of degrees of freedom (DOFs).
Scalability and high performance efficiency, specifically the efficient use of GPUs, have been the main guidelines in all the phases of the software development, from design of parallel algorithms to their implementation, as described in~\cite{BCMieee2023, BCMPDP2025}.

Sparse matrix computations represent a significant computational bottleneck in many scientific applications across various domains. They are ubiquitous in both traditional physics-based modeling and simulation, as well as in data-driven approaches such as variational data assimilation and machine learning. Specifically, preconditioned Krylov methods are the preferred approach for iteratively solving sparse linear systems in very large dimensions, as they preserve the system matrix and are therefore compatible with compressed storage schemes, thus avoiding the complications associated with the typical fill-in phenomenon encountered in direct methods. On the other hand, the low operation intensity, i.e., the low FLOP-to-byte ratio of SpMV, which is the core operation in Krylov methods, makes these methods sub optimal for current supercomputers, both in terms of performance and energy efficiency.

\texttt{BootCMatchGX} is an extensible software library available in source form\footnote{\texttt{BootCMatchGX} is available at \url{https://github.com/bootcmatch/BootCMatchGX}.}. Written in the C programming language, it employs MPI for data communication among parallel tasks and leverages the NVIDIA CUDA framework to exploit the computational power of NVIDIA GPUs. The library provides all essential functionalities for implementing several variants of the well-known Conjugate Gradient (CG) method. These include scalar products of dense vectors (dot), dense vector updates (axpy), norm computations, and SpMV, all in a distributed-memory parallel setting. At the task level, all computations are optimized for efficient execution on NVIDIA GPUs.

Sparse matrices are stored using the Compressed Sparse Row (CSR) format and are distributed across parallel tasks in blocks of contiguous rows. Special care has been taken to handle the challenges of solving systems with more than $4 \times 10^9$ degrees of freedom (DOFs) on thousands of GPUs. Using 8-byte integers for row and column indices would significantly increase memory requirements. Additionally, on GPUs, 8-byte integers introduce considerable overhead. To avoid this, the library maps global-to-local column indices using a shift mechanism. Specifically, on each GPU, the local column index is computed as the global column index (which exceeds $2^{32}-1$) minus the global index of the first row handled by that GPU. According to this convention, some column indices may temporarily become negative. However, this is only an intermediate state. Before the matrix is used, column indices are compacted and re-numbered so that all operations involve indices starting from zero. The only constraint is that the number of distinct column indices on each GPU must not exceed $2^{32}-1$, as local indices are stored in 4-byte integers. This is generally not a significant limitation, especially for sparse problems arising from PDE discretization. Importantly, there is no restriction on the number of distinct column indices in the global matrix, provided enough GPUs are available to distribute the matrix.

Communication-reduction strategies have been employed throughout the library, from the design of numerical algorithms to their implementation. These include maximizing data reuse at near-thread memory levels, minimizing host-to-GPU (and GPU-to-host) memory transfers, and overlapping GPU-level computation with inter-node communication wherever possible.

At the solver level, \texttt{BootCMatchGX} includes three distinct variants of the PCG method for solving sparse linear systems with symmetric positive-definite (SPD) coefficient matrices. These variants are: the classical algorithm originally introduced by Hestenes and Stiefel~\cite{HS1952}; a communication-reduced variant of the flexible Conjugate Gradient method, as proposed in~\cite{NN2015}; and the s-step Conjugate Gradient method developed by Chronopoulos and Gear~\cite{ChronGear1989p}. Further details on these algorithms and the implementation design patterns tailored for multi-GPU clusters can be found in~\cite{BCMieee2023,BCMPDP2025}.

The development of \texttt{BootCMatchGX} was initially motivated by the goal of providing a robust and scalable Algebraic MultiGrid (AMG) preconditioner—originally proposed in~\cite{BCMTOMS2018}—that offered improved convergence properties and performance efficiency compared to similar methods available in existing libraries, such as NVIDIA AmgX~\cite{AMGX2015}. This AMG preconditioner is based on a coarsening strategy that aggregates degrees of freedom (DOFs) using a maximum-weight matching on a weighted graph derived from the adjacency graph of the system matrix. The coarsening procedure, known as {\em Compatible weighted Matching}, has been specifically redesigned to maximize parallelism while preserving, as much as possible, the convergence properties of the preconditioner, as detailed in~\cite{BCMieee2023}.

In this work, we aim to analyze the energy efficiency profiles of the basic SpMV functionality, as well as of the main solver and preconditioner provided by \texttt{BootCMatchGX}, and compare them with similar functionalities available in state-of-the-art libraries.

\section{Energy Consumption Measurements}
\label{energy}

Measuring power consumption directly using internal or external hardware sensors is widely regarded as the most accurate method for energy assessment~\cite{bridges2016understanding}. Energy usage can be estimated by periodically sampling sensor readings during an application's execution, and the total energy consumed is then computed by integrating the power-time curve over the execution interval. Many hardware components include built-in sensors that expose power management interfaces, allowing users to monitor power usage in real time. These internal sensors are convenient, require no additional cost or hardware setup, and support fine-grained, component-level profiling. In contrast, external power measurement devices—while potentially more flexible and accurate—can be costly and impractical for large-scale or distributed systems.

In this study, we rely on internal hardware sensors to monitor CPU and GPU power consumption during execution. Our primary objective is to characterize the energy footprint of \texttt{BootCMatchGX} and other comparable state-of-the-art libraries. Rather than investigating power-saving techniques, such as DVFS, for improving GPU energy efficiency, our focus is on providing a detailed and objective assessment of the energy behavior exhibited by our application software, also in comparison with other solutions.

\subsection{CPU and GPU Power}
\label{cpuvsgpu}

On-chip power sensors integrated into modern CPU and GPU platforms provide high-frequency power measurements that are accessible through a specialized API. A prominent example is Intel’s Running Average Power Limit (RAPL) interface~\cite{khan2018rapl}. RAPL is available on Intel multicore CPUs and enables accurate monitoring of energy consumption across multiple components, including CPU cores, DRAM, and integrated GPUs. Energy usage is tracked via 32-bit Model Specific Registers (MSRs), which store cumulative energy readings since the processor was powered on. These counters are typically updated every millisecond. RAPL data can be accessed through a variety of programmatic and command-line tools, such as the Linux \texttt{sysfs} interface, performance monitoring events (\texttt{perf}), or the LIKWID tool suite~\cite{treibig2010likwid, likwid}. LIKWID is a lightweight and user-friendly set of command-line utilities and libraries designed for performance-oriented developers. It supports a range of architectures, including Intel, AMD, ARMv8, and POWER9 processors running Linux.

On the GPU side, NVIDIA GPUs are equipped with on-chip power sensors that expose power measurements via the NVIDIA Management Library (NVML) interface~\cite{nvml}. This API reports the power consumption of the GPU and its associated circuitry in milliwatts.
According to the NVML documentation, for Ampere GPUs (excluding GA100) and newer architectures, the reported values represent power averaged over a one-second interval. In contrast, for GA100 and older architectures, the API returns instantaneous power readings.

In this work, we monitor the energy consumption of CPUs and GPUs using LIKWID and NVML, respectively. 
Specifically, we employ the \texttt{lik\-wid-per\-fctr} tool from the LIKWID suite, which supports various operational modes. We use it in combination with the LIKWID MarkerAPI, a collection of functions and macros that facilitate the measurement of specific code regions. This approach allows for measuring the energy consumption of each kernel while excluding the power required for input data generation.
\texttt{likwid-perfctr} monitors the power consumption of the application from the start to the end of the execution. It generates an output file for each core used by the application, containing information on execution time, energy, and power consumption.
We do not use LIKWID for GPU monitoring; instead, we access the device directly through NVML. This approach enables us to reconstruct the GPUs’ power–time curve and more accurately estimate their static power.

For GPU power monitoring, we developed a lightweight tool called \texttt{power\-Monitor}~\cite{powerMonitor}.
The tool is built on top of \texttt{GPowerU}~\cite{GPowerU}, with both frameworks relying on the NVML interface. \texttt{GPowerU} has been developed within the TEXTAROSSA~\cite{Agostaetal2022,filgueras2024textarossa} project\footnote{\url{https://www.textarossa.eu}}, a three-year project co-funded by the European High Performance Computing (EuroHPC) JU\footnote{\url{https://eurohpc-ju.europa.eu/}}. It is a simple tool that measures the power consumption of a CUDA kernel at specific points in the device code and generates a complete power profile.
\begin{figure}[h]
\centering
\includegraphics[width=0.7\textwidth]{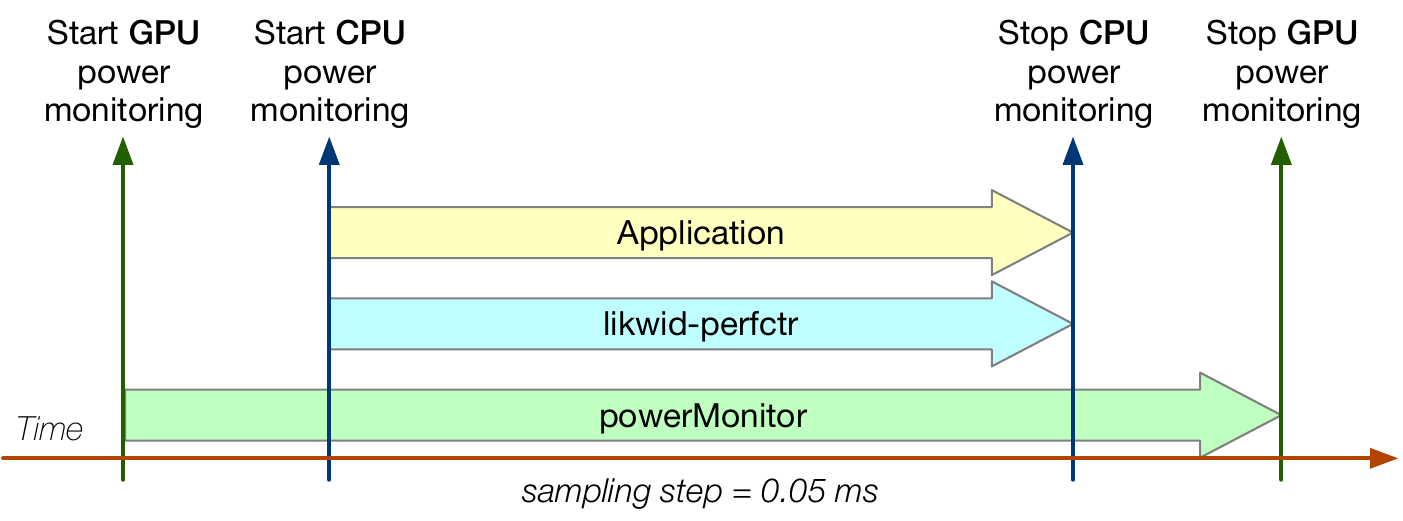}
\caption{Execution workflow for CPU and GPU power monitoring.} 
\label{fig:cpuGPUpowerMonitoring}
\end{figure}
\texttt{powerMonitor} has been adapted from \texttt{GPowerU} to better meet our specific requirements, providing a simple, yet effective, solution for monitoring the power consumption of all GPUs on a compute node. Indeed, it can be used either as an external tool or integrated directly into the application code. In our experiments, we used it as an external tool. In this case, it must be launched before the application starts and stopped once execution is complete. The complete workflow for CPU and GPU power monitoring is shown in Figure~\ref{fig:cpuGPUpowerMonitoring}. 
\texttt{powerMonitor} creates an output file for each device found on the node, containing a sequence of power samplings along with the corresponding timestamps. These samples can be used to reconstruct the power timeline of each device during the application's execution. Figure~\ref{fig:se_de_GPU} shows an example of a power–time curve. We use a sampling frequency that is higher than the actual NVML reporting resolution on A30 hardware. While this does not increase the intrinsic temporal resolution of the measurement or recover sub-second power variations, it provides a denser set of overlapping estimates of the same averaged signal. This higher sampling density yields a smoother power–time curve and reduces the discretization error when integrating power to estimate energy consumption.

Sampling inaccuracies may affect results reproducibility, particularly for workloads characterized by high-frequency GPU operations that generate short-lived power spikes. In our case study on A30 hardware, power measurements collected via the NVML API are averaged over a 1-second interval. Although this temporal aggregation enhances measurement stability, it may smooth transient fluctuations and partially mask peak power events.
In large-scale environments, concurrently polling hundreds of devices can introduce non-negligible software overhead and latency. We highlight that the proposed measurement methodology is independent of the number of cluster nodes used for the test and depends solely on the number of devices per node, as LIKWID and powerMonitor operate locally without requiring internode communication. Nevertheless, in large clusters, the accumulation of small per-device measurement errors may limit the precision of energy-efficiency comparisons across different job executions.
In our experiments, power sampling is constrained by NVML’s temporal resolution, and results are averaged over five independent runs to mitigate variability. Since the objective of this study is to provide a comparative assessment of state-of-the-art libraries rather than an exact 
quantification of absolute energy consumption, the analysis emphasizes relative energy usage in order to identify the most energy-efficient implementation.

\subsection{Static and Dynamic Energy}
\label{staticvsdynamic}

The power consumption of a device can be broadly classified into two categories: {\em static power} and {\em dynamic power}. {\em Static power} refers to the energy consumed by the device simply by being powered on, regardless of whether it is performing any computing intensive operations. {\em Dynamic power}, on the other hand, is the additional energy consumed when the device executes an application.

\begin{figure}[h]
\centering
\includegraphics[width=0.5\textwidth]{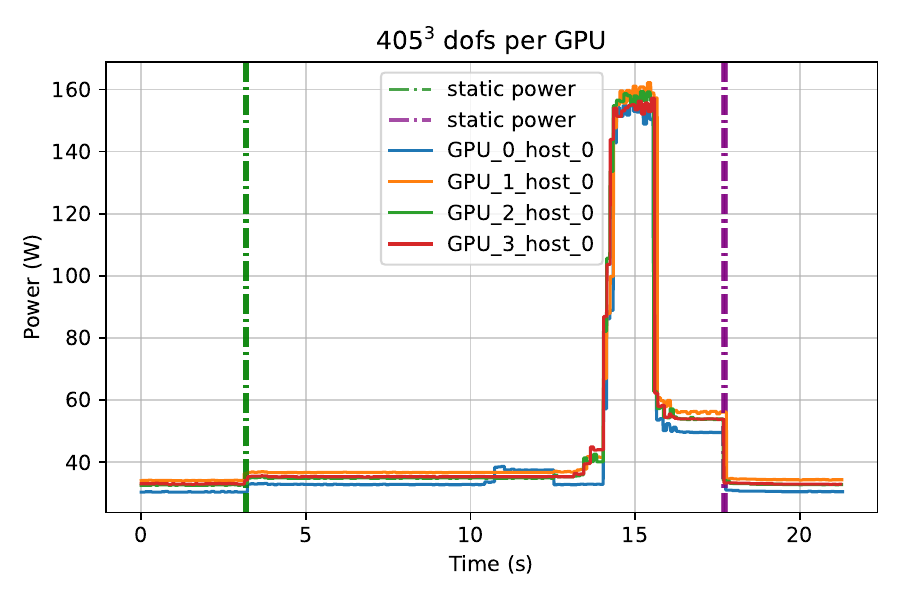}
\caption{Power–time profile of the SpMV kernel measured within the BootCMatchGX library on a single node equipped with four GPUs. The green and purple markers denote the points at which the GPUs leave and return to the idle state, respectively. These reference points are used to estimate the static power consumption of the GPUs.} 
\label{fig:se_de_GPU}
\end{figure}

Accordingly, the {\em static energy} consumed during the execution of an application is given by the product of the device's {\em static power} and the application's execution time. The {\em dynamic energy} is then calculated as the difference between the total energy consumed during execution and the {\em static energy}~\cite{lastovetsky2023energy,tan2014survey}. 
The total energy consumption ($TE$) of an application can be expressed as:
\[
TE = TP_{GPU} \times T + TP_{CPU} \times T,  \]
where $T$ is the execution time, and $TP_{GPU}$ and $TP_{CPU}$ denote the average total power for the GPU and CPU, respectively.
Similarly, the static energy ($SE$) and dynamic energy ($DE$) are defined as:
\[ SE = SP_{GPU} \times T + SP_{CPU} \times T, \]
\[ DE = TE - SE,\]
where $SP_{GPU}$ and  $SP_{CPU}$ represent the static power of the GPU and CPU, respectively.
In our analysis, we focus on the dynamic energy consumption, which corresponds to the additional energy required to execute the application.
To estimate $DE$, we proceed as follows:
\begin{itemize}
\item For the CPU, the total energy consumption $TE_{CPU}$ and the runtime $T$ are obtained using \texttt{likwid-perfctr}. The static power $SP_{CPU}$ is estimated using \texttt{likwid-powermeter} in stethoscope mode, which allows for measurements of the CPU’s power consumption over a specified time interval. We set this interval to 1 second to sample the CPU's power usage. Before running the experiments, \texttt{likwid-powermeter} is executed on idle computational nodes to measure $SP_{CPU}$, which is then multiplied by the execution time $T$ to calculate the static energy $SE_{CPU}$.
\item For the GPU, the total energy consumption $TE_{GPU}$ is computed by integrating the power timeline obtained from \texttt{powerMonitor}, which samples power approximately $20$ times per millisecond. Figure~\ref{fig:se_de_GPU} illustrates an example of the power timeline recorded during the execution of a target application on a single node with $4$ GPUs. To calculate $SE_{GPU}$, we estimate the static power 
$SP_{GPU}$ from the data and integrate it over the running time.The green and purple markers in Figure~\ref{fig:se_de_GPU} denote the transitions of the GPUs between idle and active states.
\end{itemize}
Finally, we compute the dynamic energies $DE_{GPU}$ and $DE_{CPU}$, and sum them to obtain the total dynamic energy consumption: 
\[ DE = DE_{GPU} + DE_{CPU}.
\]

\section{Results and Comparisons}
\label{results}

In this section, we analyze the performance and energy consumption of \texttt{BootCMatchGX}, comparing it with state-of-the-art libraries. Tests were conducted on a cluster with dual-socket Intel Xeon Gold CPUs (32 cores each) and four NVIDIA A30 GPUs per node, interconnected via HDR InfiniBand; up to 16 nodes (64 GPUs) were used. The software stack included CUDA 12.3, Open MPI 4.1.6, and GCC 12.2.1.

We evaluate the SpMV operation, as a fundamental sparse matrix operation, and of the CG solver. \texttt{BootCMatchGX} (v1.1.0) is compared with Ginkgo (v1.9.0)~\cite{ginkgo, ginkgo-toms-2022} for SpMV and un-precon\-di\-tioned CG, while preconditioned CG results are also compared with NVIDIA AmgX (v2.4.0)~\cite{naumov2015amgx,AmgX(2025)}. The AmgX SpMV implementation is not considered, as no official driver is available to test it separately.

Benchmark problems are derived from the 3D Poisson equation with homogeneous Dirichlet boundary conditions, discretized on uniform meshes with 7- and 27-point stencils (the latter as in HPCG~\cite{HPCG}). We analyze performance and energy efficiency under both strong and weak scalability conditions. In both cases, the local problem size is initially chosen as the largest that fits into the memory capacity of a single GPU, ensuring full exploitation of device resources and realistic memory-bound conditions. In strong scaling experiments, this global problem (corresponding to the single-GPU memory-saturating size) is kept fixed and partitioned among GPUs as their number increases, leading to a reduced local workload per GPU. In weak scaling experiments, instead, the global problem size grows linearly with the number of GPUs, while the local workload per GPU remains constant at the memory-saturating size. Matrices are partitioned by rows across GPUs, with the 3D domain mapped to a 3D grid of MPI tasks, reproducing realistic communication/computation patterns typical of large-scale PDE-based applications.

In addition to Poisson-related matrices, we consider a set of five symmetric positive definite sparse matrices selected from the \emph{SuiteSparse Matrix Collection}~\cite{SuiteSparseCollection}. The matrices are chosen among SPD problems that do not necessarily arise from the discretization of diffusion-dominated equations on structured grids, thereby providing test cases with diverse sparsity patterns and numerical features that often pose significant challenges for Krylov solvers coupled with AMG preconditioners.
Key characteristics of the selected matrices are summarized in Table~\ref{tab:suitesparse_matrices}. Since these problems are defined at a fixed size, the analysis is restricted to \emph{strong scaling} experiments. Although the selected matrices exhibit relatively large dimensions compared to other entries in the collection, their overall size makes it reasonable to limit the evaluation to a small number of accelerators. Accordingly, experiments are performed using up to 4 GPUs.

\begin{table}[t]
\centering
\caption{Key characteristics of the selected SPD matrices from the SuiteSparse Matrix Collection.}
\label{tab:suitesparse_matrices}
\begin{tabular}{lrrr}
\hline
Matrix & Rows & Nonzeros & Avg. nnz/row \\
\hline
\texttt{G3\_circuit}         & 1\,585\,478 & 7\,660\,826  & 4.8  \\
\texttt{af\_shell8}  & 504\,855   & 17\,579\,155 & 34.8 \\
\texttt{boneS10}      & 914\,898   & 40\,878\,708 & 44.7 \\
\texttt{ecology2}          & 999\,999   & 4\,995\,991  & 5.0  \\
\texttt{parabolic\_fem} & 525\,825   & 3\,674\,625  & 7.0  \\
\hline
\end{tabular}
\end{table}

\subsection{Results on Poisson-related benchmarks}
\label{res:poisson}

\subsubsection{SpMV Operation}
\label{spmv}

We begin by analyzing the performance behavior of the SpMV operation under both strong and weak scalability scenarios.For the single-GPU case, the problem size is set to $405^3$ DOFs for the 7-point stencil and $260^3$ DOFs for the 27-point stencil, corresponding to the largest dimensions that fully saturate the available GPU memory. All performance measurements reported in the figures represent averages over five independent runs, with each run consisting of 100 repetitions of the SpMV computation. All plots include 95\% confidence intervals, shown as black vertical lines. 
\begin{figure}[h]
\centering
     \begin{subfigure}[t]{0.48\textwidth}
         \centering
         \includegraphics[width=\textwidth]{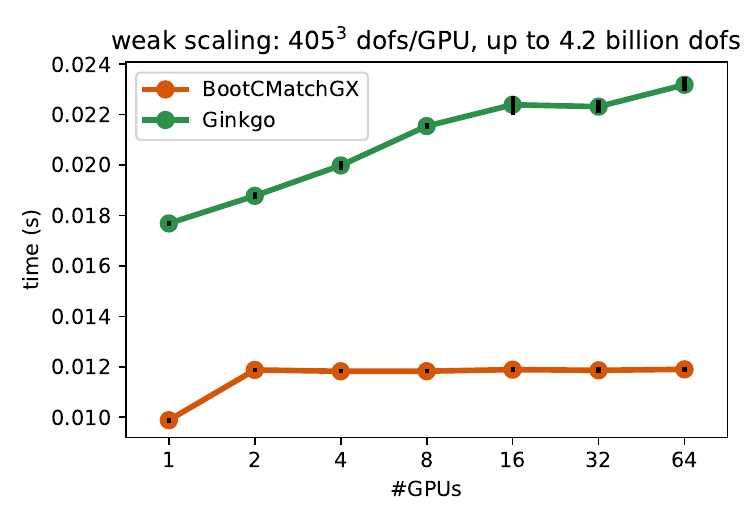}
         \caption{7-points stencil matrix with $405^3$ DOFs per GPU under weak scalability.}
         \label{fig:spmv_timeWeak7}
     \end{subfigure}
     \hfill
     \begin{subfigure}[t]{0.48\textwidth}
         \centering
         \includegraphics[width=\textwidth]{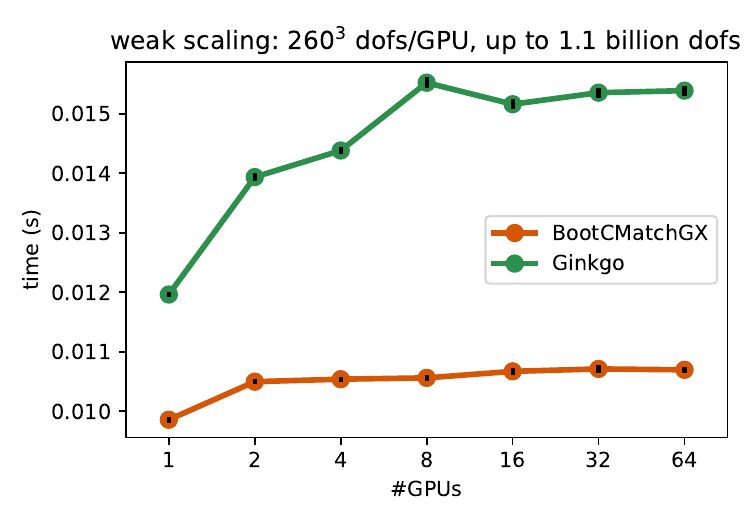}
         \caption{27-points stencil matrix with $260^3$ DOFs per GPU under weak scalability.}
         \label{fig:spmv_timeWeak27}
     \end{subfigure}
    \begin{subfigure}{0.48\textwidth}
         \centering
         \includegraphics[width=\textwidth]{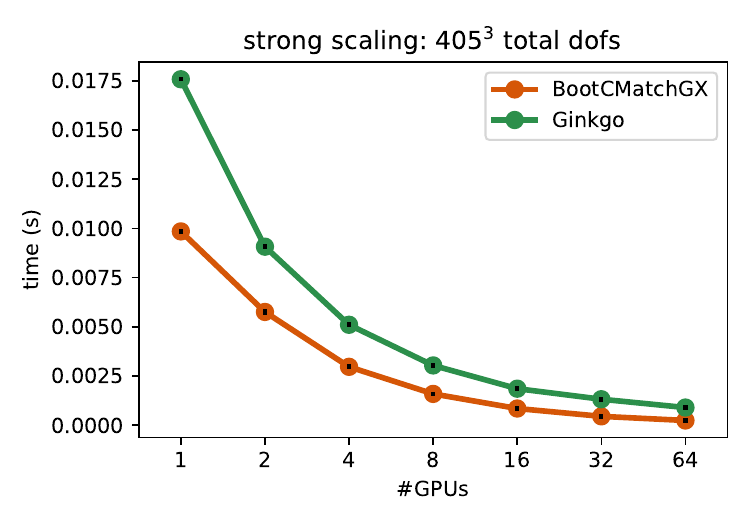}
         \caption{7-points stencil matrix with a total of  $405^3$ DOFs under strong scalability.}
         \label{fig:spmv_timeStrong7}
     \end{subfigure}
     \hfill
     \begin{subfigure}{0.48\textwidth}
         \centering
         \includegraphics[width=\textwidth]{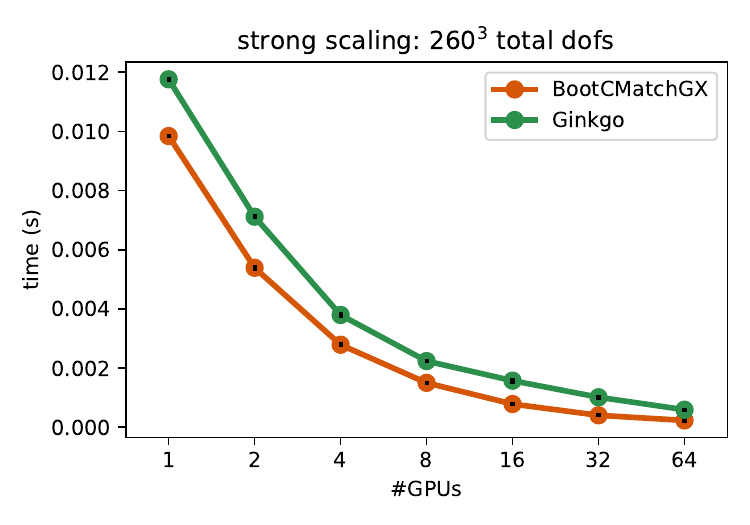}
         \caption{27-points stencil matrix with a total of $260^3$ DOFs under strong scalability.}
         \label{fig:spmv_timeStrong27}
     \end{subfigure}
\caption{SpMV execution times under weak and strong scalability scenarios.}
\label{fig:spmv_time}
\end{figure}

Figure~\ref{fig:spmv_time} shows the execution times of the SpMV computation, comparing the \texttt{BootCMatchGX} and Ginkgo implementations. Across all tested scenarios, \texttt{BootCMatchGX} consistently outperforms Ginkgo, with notably lower execution times. The performance gap is especially pronounced in the weak scalability cases. These results suggest a potential advantage for \texttt{BootCMatchGX} in terms of both performance and energy efficiency for very large-scale computations.

\begin{figure}[h]
\centering
     \begin{subfigure}[t]{0.48\textwidth}
         \centering
         \includegraphics[width=\textwidth]{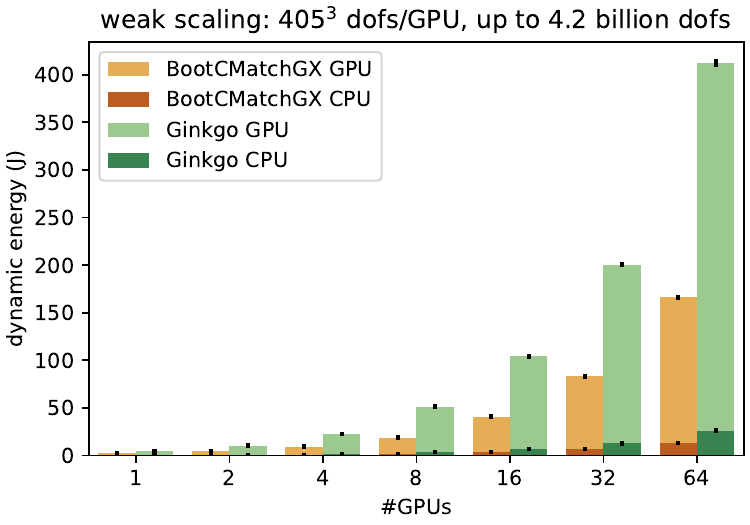}
         \caption{7-points stencil matrix with $405^3$ DOFs per GPU under weak scalability.}
         \label{fig:spmv_dynamic_energy_weak_7}
     \end{subfigure}
     \hfill
     \begin{subfigure}[t]{0.48\textwidth}
         \centering
         \includegraphics[width=\textwidth]{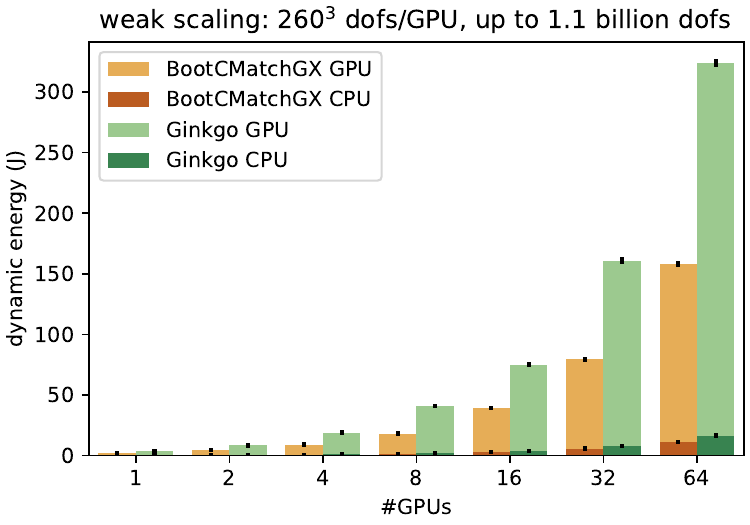}
         \caption{27-points stencil matrix with $260^3$ DOFs per GPU under weak scalability.}
         \label{fig:spmv_dynamic_energy_weak_27}
     \end{subfigure}
    \begin{subfigure}[t]{0.48\textwidth}
         \centering
         \includegraphics[width=\textwidth]{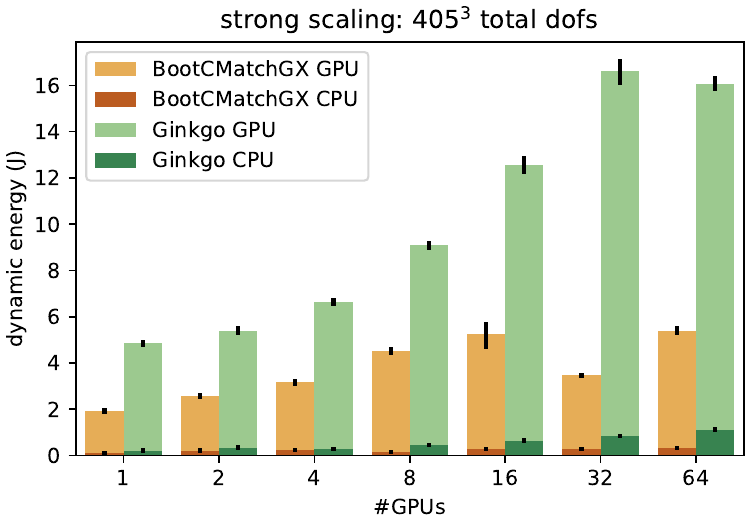}
         \caption{7-points stencil matrix with a total of $405^3$ DOFs under strong scalability.}
         \label{fig:spmv_dynamic_energy_strong_7}
     \end{subfigure}
     \hfill
     \begin{subfigure}[t]{0.48\textwidth}
         \centering
         \includegraphics[width=\textwidth]{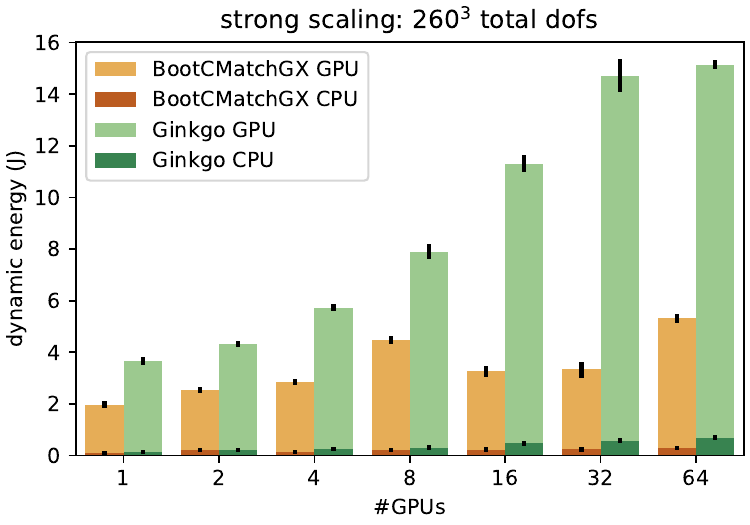}
         \caption{27-points stencil matrix with a total of $260^3$ DOFs under strong scalability.}
         \label{fig:spmv_dynamic_energy_strong_27}
     \end{subfigure}
\caption{Dynamic energy consumption breakdown of the SpMV computation on GPU and CPU under weak and strong scalability scenarios.} 
\label{fig:spmv_dynamic_energy}
\end{figure}
Figure~\ref{fig:spmv_dynamic_energy} reports the breakdown of dynamic energy consumption for the SpMV computation, with GPU and CPU contributions represented as distinct colored segments in each bar, as indicated in the legend. The results demonstrate that the BootCMatchGX implementation is consistently more energy-efficient than Ginkgo under both weak and strong scalability conditions. In all cases, its dynamic energy consumption is approximately half that of Ginkgo. The data also reveal that the CPU contribution to the overall energy consumption is negligible compared to that of the GPU. This outcome is expected, as the bulk of the computational workload is offloaded to the GPU, while the CPU is primarily responsible for inter-process communication—a task that requires limited computational effort and therefore incurs only a marginal energy cost.

\begin{figure}[h]
\centering
     \begin{subfigure}[t]{0.48\textwidth}
         \centering
         \includegraphics[width=\textwidth]{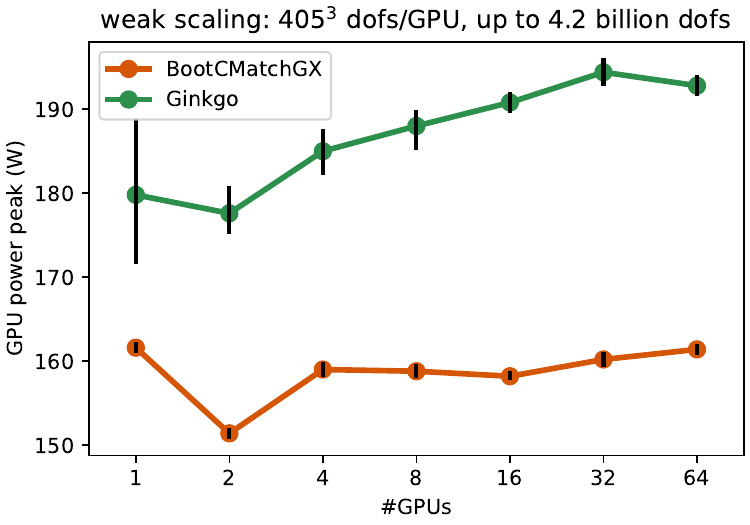}
         \caption{7-points stencil matrix with $405^3$ DOFs per GPU under weak scalability.}
         \label{fig:spmv_power_peak_weak_7}
     \end{subfigure}
     \hfill
     \begin{subfigure}[t]{0.48\textwidth}
         \centering
         \includegraphics[width=\textwidth]{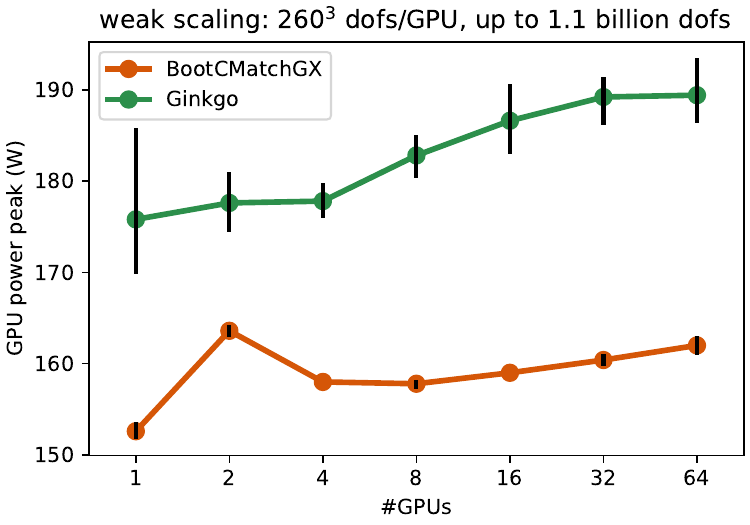}
         \caption{27-points stencil matrix with $260^3$ DOFs per GPU under weak scalability.}
         \label{fig:spmv_power_peak_weak_27}
     \end{subfigure}
    \begin{subfigure}[t]{0.48\textwidth}
         \centering
         \includegraphics[width=\textwidth]{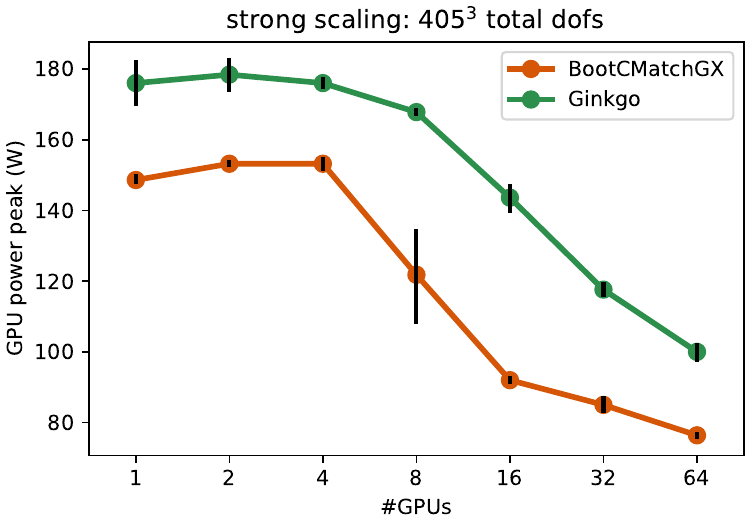}
         \caption{7-points stencil matrix with a total of $405^3$ DOFs under strong scalability.}
         \label{fig:spmv_power_peak_strong_7}
     \end{subfigure}
     \hfill
     \begin{subfigure}[t]{0.48\textwidth}
         \centering
         \includegraphics[width=\textwidth]{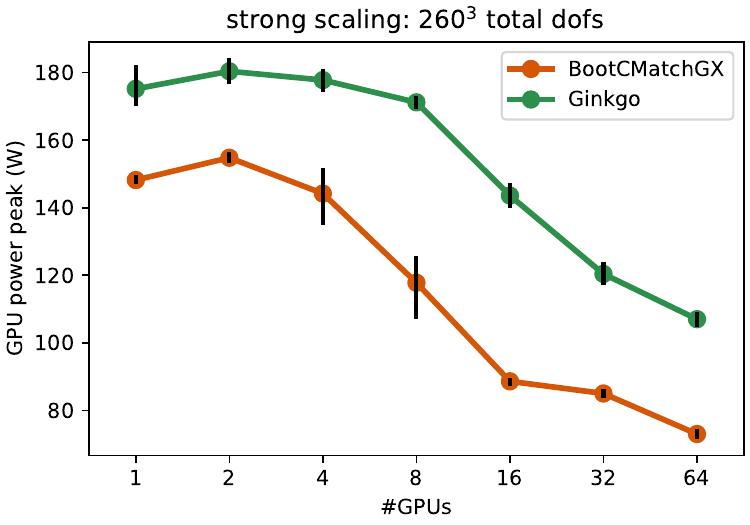}
         \caption{27-points stencil matrix with a total of $260^3$ DOFs under strong scalability.}
         \label{fig:spmv_power_peak_strong_27}
     \end{subfigure}
\caption{GPU power peak of the SpMV computation under weak and strong scalability scenarios.} 
\label{fig:spmv_power_energy}
\end{figure}
Figure~\ref{fig:spmv_power_energy} shows the GPU power peaks recorded during the execution of the SpMV computation. As the charts clearly illustrate, \texttt{BootCMatchGX} consistently exhibits lower GPU power peaks than Ginkgo, indicating a more efficient use of GPU resources by maintaining a steady workload and avoiding sudden power spikes. As expected, under strong scalability conditions, the power peaks decrease with an increasing number of GPUs, due to the reduced workload per GPU.

\begin{figure}[h]
\centering
     \begin{subfigure}[t]{0.48\textwidth}
         \centering
         \includegraphics[width=\textwidth]{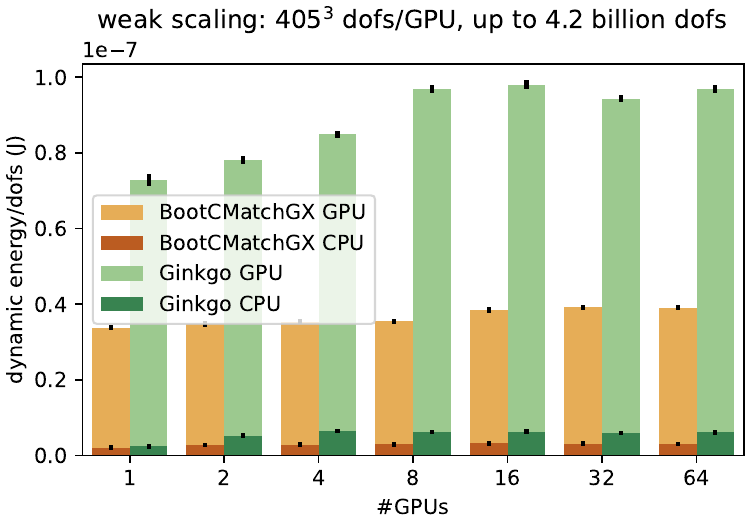}
         \caption{7-points stencil matrix with $405^3$ DOFs under weak scalability.}
         \label{fig:spmv_dofs_weak_7}
     \end{subfigure}
     \hfill
     \begin{subfigure}[t]{0.48\textwidth}
         \centering
         \includegraphics[width=\textwidth]{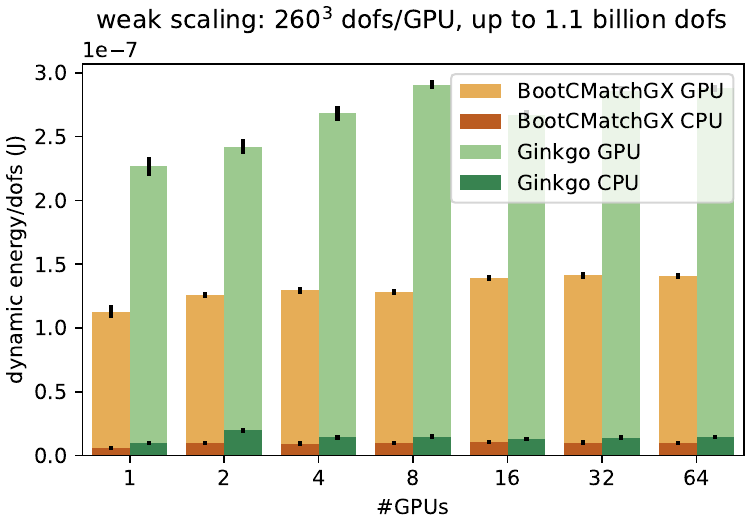}
         \caption{27-points stencil matrix with $260^3$ DOFs under weak scalability.}
         \label{fig:spmv_dofs_weak_27}
     \end{subfigure}
    \begin{subfigure}[t]{0.48\textwidth}
         \centering
         \includegraphics[width=\textwidth]{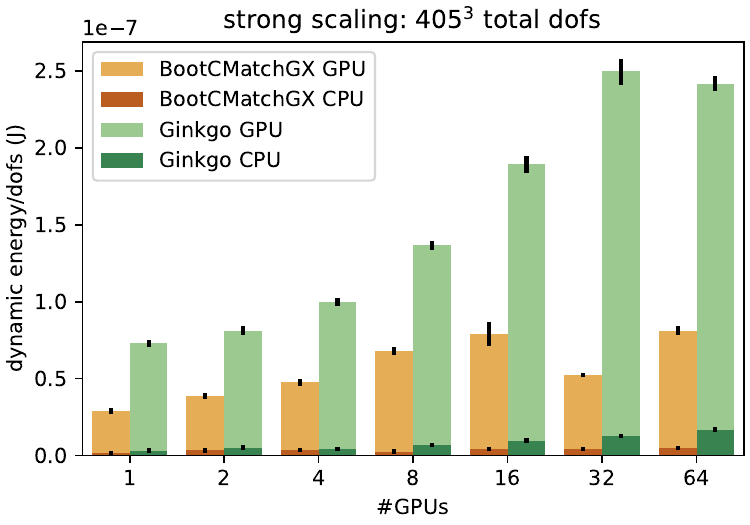}
         \caption{7-points stencil matrix with a total of $405^3$ DOFs under strong scalability.}
         \label{fig:spmv_dofs_strong_7}
     \end{subfigure}
     \hfill
     \begin{subfigure}[t]{0.48\textwidth}
         \centering
         \includegraphics[width=\textwidth]{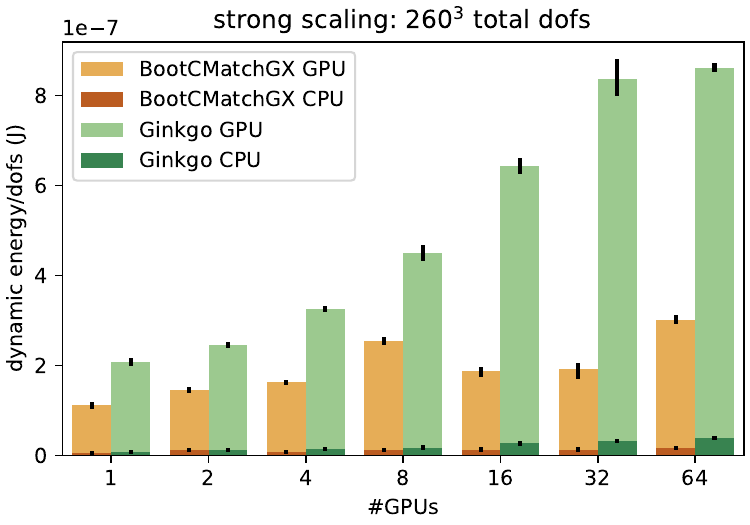}
         \caption{27-points stencil matrix with a total of $260^3$ DOFs under strong scalability.}
         \label{fig:spmv_dofs_strong_27}
     \end{subfigure}
\caption{Dynamic energy consumption per DOF breakdown of the SpMV computation under weak and strong scalability scenarios.} 
\label{fig:spmv_dofs}
\end{figure}
Figure~\ref{fig:spmv_dofs} presents the dynamic energy consumption per DOFs. As anticipated from previous results, \texttt{BootCMatchGX} demonstrates significantly higher energy efficiency than Ginkgo—approximately twice as efficient. In both implementations, energy efficiency remains nearly constant with increasing numbers of GPUs under weak scalability, confirming the good scalability of both solutions. Under strong scalability, as expected, efficiency declines as the number of GPUs increases, due to the reduction of the computational workload per GPU. 

Finally, Tables~\ref{tab:spmv_dynamicVSstatic_7p} and~\ref{tab:spmv_dynamicVSstatic_27p} compare static and dynamic energy consumption, expressing the latter as a percentage of the static energy. Although \texttt{BootCMatchGX} shows a higher dynamic energy percentage, this should be interpreted in light of its design strategy. In weak scalability scenarios, the decomposition of static and dynamic energy provides insight into sustainability trade-offs. Since static energy is time-dependent, reductions in communication overhead and idle phases — achieved through improved data locality and communication-aware design — directly limit its contribution.
Therefore, the higher dynamic energy share observed for \texttt{BootCMatchGX} likely reflects a larger fraction of energy devoted to useful computation rather than baseline dissipation. When considered together with its shorter runtimes and lower total dynamic energy consumption, this behavior supports the view that \texttt{BootCMatchGX} achieves more effective GPU utilization, with a lower overall energy consumption.

\begin{table}
\centering
\caption{Static vs. dynamic energy consumption, for the 7-points stencil matrix with $405^3$ DOFs under weak scalability. Columns report GPU, CPU, and total dynamic energy consumption expressed as a percentage of the static energy.}
\label{tab:spmv_dynamicVSstatic_7p}
\begin{tabular}{llrrr}
\toprule
 &  & GPU \% & CPU \% & total \% \\
\#GPUs & library &  &  &  \\
\midrule
\multirow[t]{2}{*}{1} & BCMGX & 47.29 & 16.12 & 42.25 \\
 & Ginkgo & 39.77 & 10.45 & 36.38 \\
\cline{1-5}
\multirow[t]{2}{*}{2} & BCMGX & 63.94 & 17.61 & 52.94 \\
 & Ginkgo & 39.52 & 21.33 & 37.38 \\
\cline{1-5}
\multirow[t]{2}{*}{4} & BCMGX & 61.75 & 18.83 & 51.99 \\
 & Ginkgo & 42.03 & 24.44 & 39.86 \\
\cline{1-5}
\multirow[t]{2}{*}{8} & BCMGX & 61.91 & 19.00 & 52.15 \\
 & Ginkgo & 38.54 & 22.04 & 36.77 \\
\cline{1-5}
\multirow[t]{2}{*}{16} & BCMGX & 51.74 & 20.61 & 45.96 \\
 & Ginkgo & 38.99 & 21.42 & 37.03 \\
\cline{1-5}
\multirow[t]{2}{*}{32} & BCMGX & 60.05 & 20.41 & 51.89 \\
 & Ginkgo & 43.06 & 20.42 & 40.23 \\
\cline{1-5}
\multirow[t]{2}{*}{64} & BCMGX & 51.71 & 19.82 & 45.88 \\
 & Ginkgo & 40.78 & 20.08 & 38.28 \\
\cline{1-5}
\bottomrule
\end{tabular}
\end{table}

\begin{table}
\centering
\caption{Static vs. dynamic energy consumption, for the 27-points stencil matrix with a total of $260^3$ DOFs under weak scalability. Columns report GPU, CPU, and total dynamic energy consumption expressed as a percentage of the static energy. }
\label{tab:spmv_dynamicVSstatic_27p}
\begin{tabular}{llrrr}
\toprule
 &  & GPU \% & CPU \% & total \% \\
\#GPUs & library &  &  &  \\
\midrule
\multirow[t]{2}{*}{1} & BCMGX & 36.75 & 11.81 & 33.16 \\
 & Ginkgo & 28.95 & 16.58 & 28.04 \\
\cline{1-5}
\multirow[t]{2}{*}{2} & BCMGX & 45.68 & 18.81 & 41.12 \\
 & Ginkgo & 35.58 & 28.53 & 34.88 \\
\cline{1-5}
\multirow[t]{2}{*}{4} & BCMGX & 46.27 & 18.08 & 41.56 \\
 & Ginkgo & 37.23 & 20.14 & 35.61 \\
\cline{1-5}
\multirow[t]{2}{*}{8} & BCMGX & 47.12 & 18.68 & 42.23 \\
 & Ginkgo & 34.73 & 19.31 & 33.37 \\
\cline{1-5}
\multirow[t]{2}{*}{16} & BCMGX & 40.56 & 20.28 & 37.66 \\
 & Ginkgo & 39.44 & 17.52 & 37.15 \\
\cline{1-5}
\multirow[t]{2}{*}{32} & BCMGX & 41.20 & 19.04 & 38.03 \\
 & Ginkgo & 36.77 & 18.45 & 35.05 \\
\cline{1-5}
\multirow[t]{2}{*}{64} & BCMGX & 42.25 & 18.56 & 38.79 \\
 & Ginkgo & 37.08 & 19.10 & 35.38 \\
\cline{1-5}
\bottomrule
\end{tabular}
\end{table}

Overall, our analysis shows that the \texttt{BootCMatchGX} implementation of SpMV outperforms Ginkgo in both execution time and energy consumption. Its improved performance directly contributes to a lower energy footprint. Moreover, the consistently lower GPU power peaks observed for \texttt{BootCMatchGX} (see Figure~\ref{fig:spmv_power_energy}) suggest a more effective utilization of GPU resources, further reinforcing its energy-efficiency features.

\begin{figure}[ht!]
\centering
     \begin{subfigure}[t]{0.48\textwidth}
         \centering
         \includegraphics[width=\textwidth]{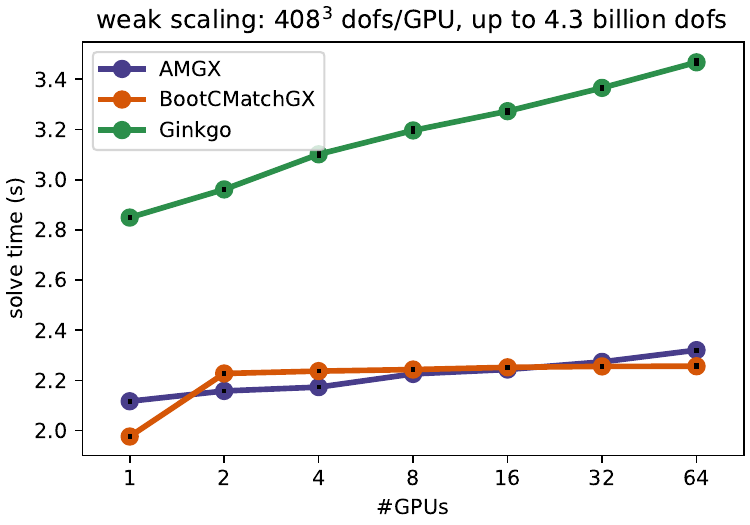}
         \caption{7-points stencil matrix with $408^3$ DOFs per GPU under weak scalability.}
         \label{fig:cg_timeWeak7}
     \end{subfigure}
     \hfill
     \begin{subfigure}[t]{0.48\textwidth}
         \centering
         \includegraphics[width=\textwidth]{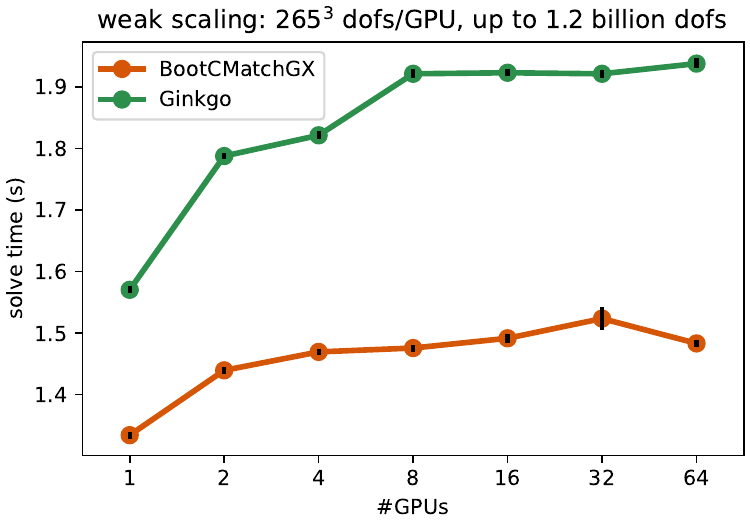}
         \caption{27-points stencil matrix with $265^3$ DOFs per GPU under weak scalability.}
         \label{fig:cg_timeWeak27}
     \end{subfigure}
     \hfill
     \begin{subfigure}[t]{0.48\textwidth}
         \centering
         \includegraphics[width=\textwidth]{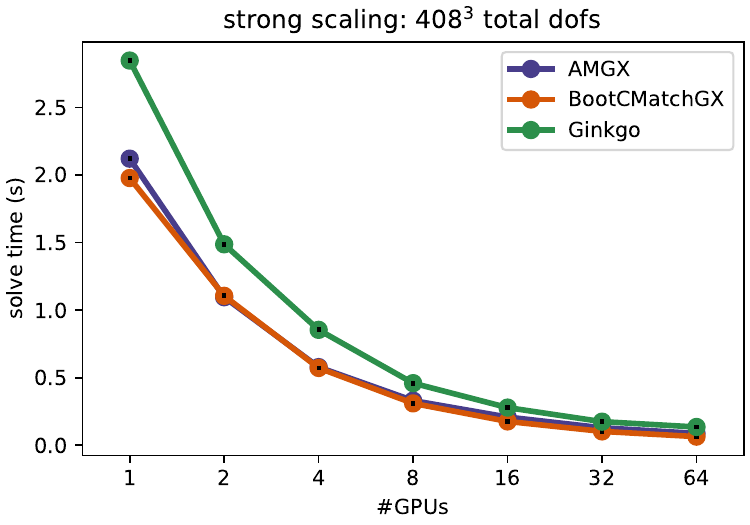}
         \caption{7-points stencil matrix with a total of $408^3$ DOFs under strong scalability.}
         \label{fig:cg_timeStrong7}
     \end{subfigure}
     \hfill
     \begin{subfigure}[t]{0.48\textwidth}
         \centering
         \includegraphics[width=\textwidth]{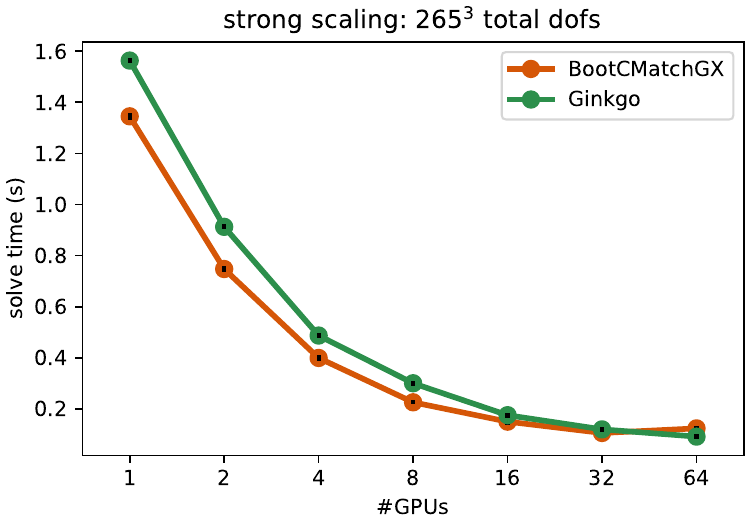}
         \caption{27-points stencil matrix with a total of $265^3$ DOFs under strong scalability.}
         \label{fig:cg_timeStrong27}
     \end{subfigure}
\caption{Un-preconditioned CG execution times under weak and strong scalability scenarios.}
\label{fig:cg_time}
\end{figure}
\subsubsection{Un-preconditioned Conjugate Gradient solver}
\label{cg}

We evaluate the un-preconditioned CG implementations of \texttt{BootCMatchGX}, Ginkgo, and NVIDIA AmgX under both strong and weak scalability. The analysis of the un-preconditioned solver is of particular interest, as it provides a direct insight on the intrinsic computational costs of a CG iteration without the influence of preconditioning. In this setting, the performance can be directly attributed to the efficiency of the SpMV operation—by far the dominant kernel in terms of execution time—as well as to the relative impact of other fundamental operations such as axpy, dot products, and global reductions. This perspective is essential for isolating the contribution of these core building blocks and for assessing their impact independently of any preconditioning strategy. 

For the single-GPU case, the problem size is set to $408^3$ DOFs with the 7-point stencil and $265^3$ DOFs with the 27-point stencil; AmgX is excluded from the 27-point case, as this benchmark is not supported. Reported results are averages over five runs and include 95\% confidence intervals, shown as black vertical lines. For some configurations (1, 32, and 64 GPUs) of the 7-point stencil in the weak scaling settings, we ran also the {\em t-test}.  The result of the t-test confirms that the null hypotheses (the AmgX, BootCMatchGX, and Ginkgo samples have the same mean) is always rejected. As a consequence, we are confident that the differences are not due to noise in the measures.
For our tests, we set the maximum number of iterations to $100$ and the relative residual tolerance to $10^{-16}$. Since our focus is on the cost per iteration rather than on convergence properties, this setup ensures that each implementation performs exactly 100 iterations in all scenarios.

Figure~\ref{fig:cg_time} reports the execution times of the CG solver. Across all scenarios, \texttt{BootCMatchGX} consistently outperforms Ginkgo, achieving substantially lower runtimes, particularly under weak scalability. For the 7-point stencil, NVIDIA AmgX also shows superior performance compared to Ginkgo, while delivering results comparable to those of \texttt{BootCMatchGX}. In strong scalability tests, however, the gap narrows as the number of GPUs increases, possibly due to the growing impact of communication and synchronization overheads. Under these conditions, possible advantage of specialized GPU implementations diminishes, and Ginkgo achieves performance comparable to both \texttt{BootCMatchGX} and AmgX.
However, we can conclude that both NVIDIA AmgX and \texttt{BootCMatchGX} show clear advantages over Ginkgo in terms of performance, and potentially energy efficiency, for large-scale computations. 

\begin{figure}[h]
\centering
     \begin{subfigure}[t]{0.48\textwidth}
         \centering
         \includegraphics[width=\textwidth]{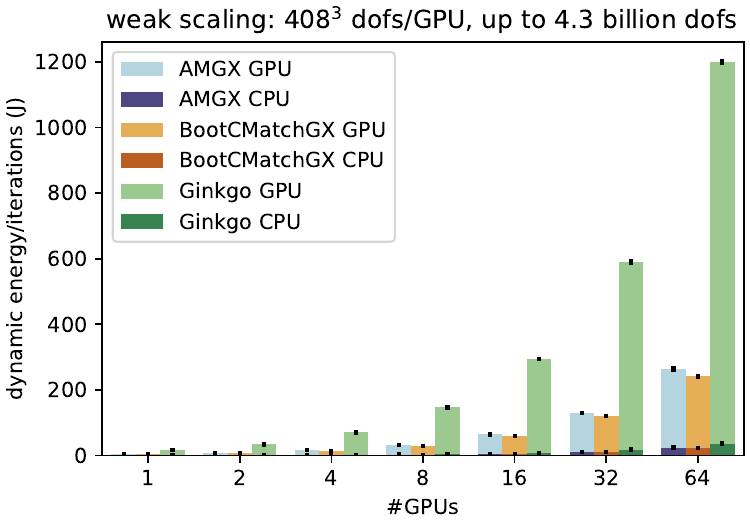}
         \caption{7-points stencil matrix with $408^3$ DOFs per GPU under weak scalability.}
         \label{fig:cg_energyIterWeak7}
     \end{subfigure}
     \hfill
    \begin{subfigure}[t]{0.48\textwidth}
         \centering
         \includegraphics[width=\textwidth]{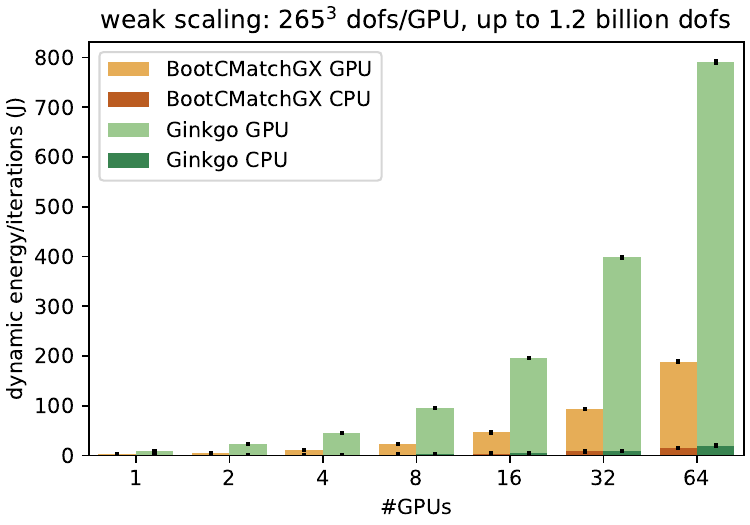}
         \caption{27-points stencil matrix with $265^3$ DOFs per GPU under weak scalability.}
         \label{fig:cg_energyIterWeak27}
     \end{subfigure}
     \hfill
     \begin{subfigure}[t]{0.48\textwidth}
         \centering
         \includegraphics[width=\textwidth]{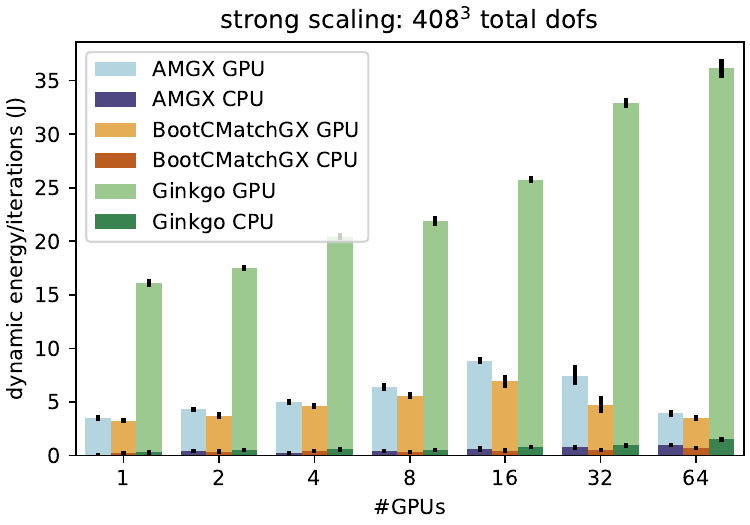}
         \caption{7-points stencil matrix with a total of $408^3$ DOFs under strong scalability.}
         \label{fig:cg_energyIterStrong7}
     \end{subfigure}
     \hfill
     \begin{subfigure}[t]{0.48\textwidth}
         \centering
         \includegraphics[width=\textwidth]{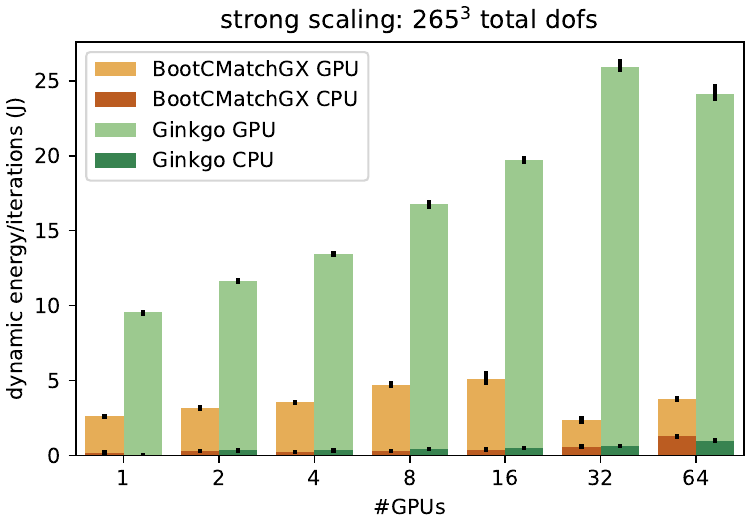}
         \caption{27-points stencil matrix with a total of $265^3$ DOFs under strong scalability.}
         \label{fig:cg_energyIterStrong27}
     \end{subfigure}
\caption{Dynamic energy consumption per iteration breakdown of the un-preconditioned CG computation on GPU and CPU under weak and strong scalability scenarios.} 
\label{fig:cg_energyIter}
\end{figure}
Figure~\ref{fig:cg_energyIter} breaks down the dynamic energy consumption per iteration for the CG computation, with GPU and CPU contributions shown as colored segments in each bar (see legend). 
The results indicate that both NVIDIA AmgX and \texttt{BootCMatchGX} are consistently more energy-efficient than Ginkgo under both weak and strong scalability conditions. Their per-iteration dynamic energy consumption is less than half of Ginkgo, while AmgX and \texttt{BootCMatchGX} exhibit comparable overall energy usage. Notably, \texttt{BootCMatchGX} achieves a slight advantage over AmgX in both weak and strong scalability cases.

\begin{figure}[h]
\centering
     \begin{subfigure}[t]{0.48\textwidth}
         \centering
         \includegraphics[width=\textwidth]{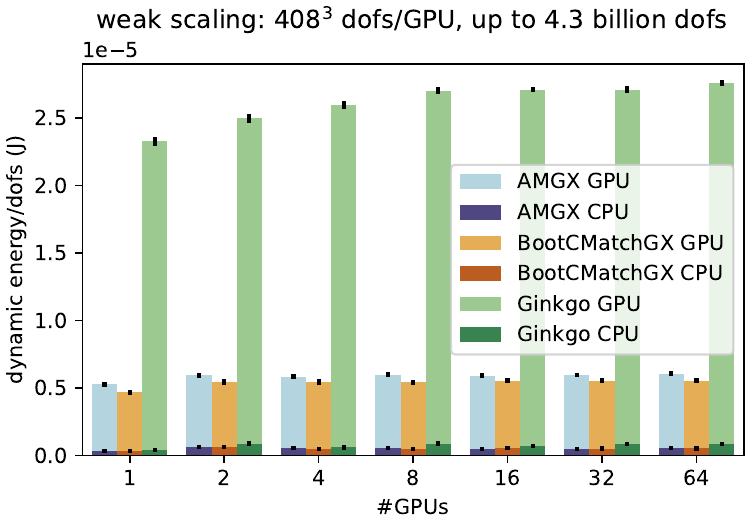}
         \caption{7-points stencil matrix with $408^3$ DOFs per GPU under weak scalability.}
         \label{fig:cg_energyDofsWeak7}
     \end{subfigure}
     \hfill
    \begin{subfigure}[t]{0.48\textwidth}
         \centering
         \includegraphics[width=\textwidth]{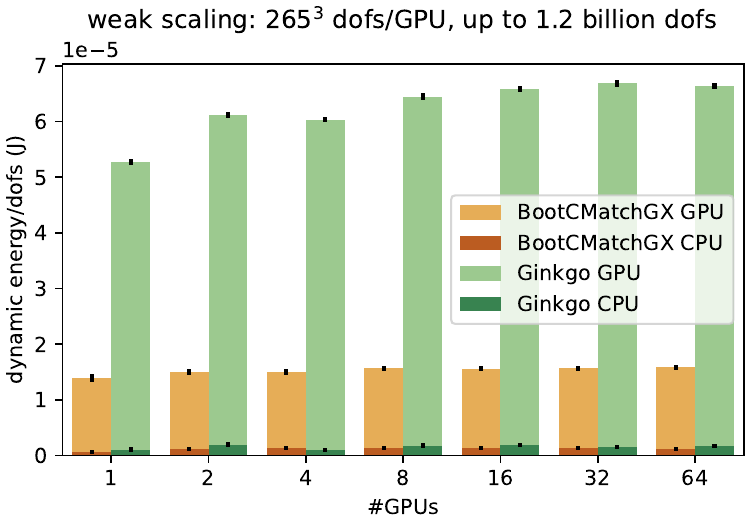}
         \caption{27-points stencil matrix with $265^3$ DOFs per GPU under weak scalability.}
         \label{fig:cg_energyDofsWeak27}
     \end{subfigure}
     \hfill
     \begin{subfigure}[t]{0.48\textwidth}
         \centering
         \includegraphics[width=\textwidth]{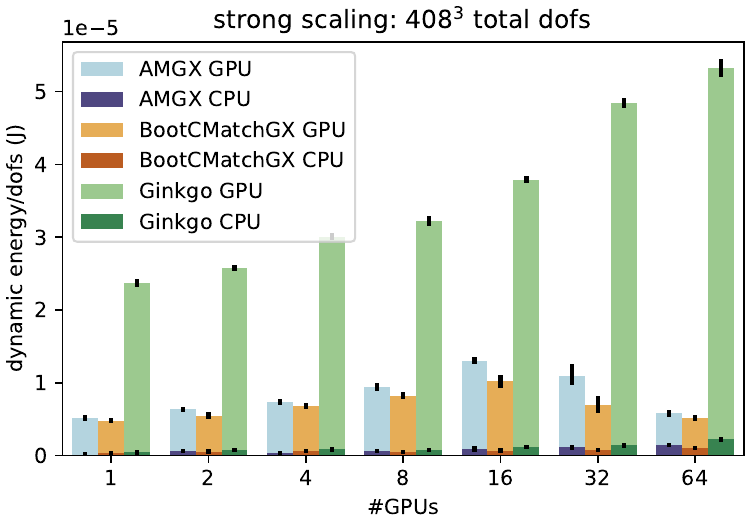}
         \caption{7-points stencil matrix with a total of $408^3$ DOFs under strong scalability.}
         \label{fig:cg_energyDofsStrong7}
     \end{subfigure}
     \hfill
     \begin{subfigure}[t]{0.48\textwidth}
         \centering
         \includegraphics[width=\textwidth]{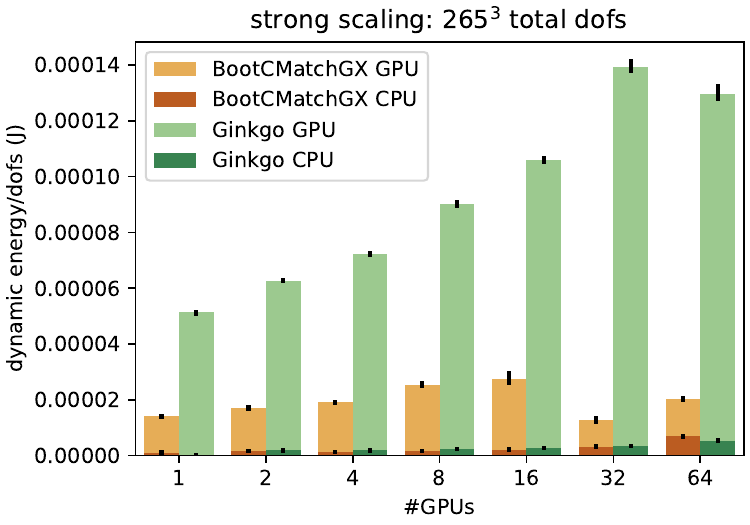}
         \caption{27-points stencil matrix with a total of $265^3$ DOFs under strong scalability.}
         \label{fig:cg_energyDofsStrong27}
     \end{subfigure}
\caption{Dynamic energy consumption per DOF breakdown of the un-preconditioned CG computation on GPU and CPU under weak and strong scalability scenarios.} 
\label{fig:cg_energyDofs}
\end{figure}
Figure~\ref{fig:cg_energyDofs} breaks down the dynamic energy consumption per DOF for the CG computation, with GPU and CPU contributions shown as colored segments in each bar (see legend). Under weak scalability, all implementations exhibit stable per-DOF energy consumption as the problem size increases. In other words, the energy required per DOF remains nearly constant when the number of processes grows proportionally to the global problem size, indicating that the additional computational resources are effectively amortized and that none of the implementations introduces significant energy overhead as the scale increases.
The energy profile of the un-preconditioned CG solver follows the same relative trends observed for the SpMV operation in \ref{spmv}. This is expected, since SpMV is by far the dominant kernel within each iteration, and the improvements achieved by BootCMatchGX over Ginkgo in the standalone SpMV directly carry over to the full solver. In absolute terms, however, the dynamic energy per DOF in CG is about two orders of magnitude higher than in SpMV, reflecting the cumulative cost of performing many iterations rather than a single matrix–vector product. Although additional vector operations are executed in each iteration, their contribution does not alter the overall picture, which remains dictated by the efficiency of the SpMV kernel. This explains why BootCMatchGX (and, in the 7-point stencil case, NVIDIA AmgX) consistently achieve significantly lower dynamic energy consumption per iteration and per DOF compared to Ginkgo, as confirmed by the results in Figures \ref{fig:cg_energyIter} and \ref{fig:cg_energyDofs}. In summary, as expected, the comparative trends observed in CG largely reflect those already highlighted for SpMV, underlining the central role of the sparse matrix–vector product in defining the solver’s energy efficiency.

Tables~\ref{tab:solver_dynamicVSstatic_noprec7p} and~\ref{tab:solver_dynamicVSstatic_noprec27p} compare static and dynamic energy consumption, reporting the latter as a percentage of the static energy. Ginkgo consistently exhibits higher dynamic energy percentages than \texttt{BootCMatchGX}, and its dynamic energy consumption and runtimes are also greater, indicating less efficient GPU utilization and, consequently, higher total energy consumption.
AmgX shows higher dynamic energy percentages than both Ginkgo and \texttt{BootCMatchGX}; however, its dynamic energy consumption and runtime are lower than Ginkgo’s, leading to lower overall energy consumption as a result of more effective GPU utilization.
Although \texttt{BootCMatchGX} and AmgX achieve comparable runtimes, \texttt{BootCMatchGX} attains lower dynamic energy consumption—and therefore lower total energy consumption—indicating more efficient GPU resource utilization associated with a lower average power draw.

\begin{table}
\centering
\caption{Static vs. dynamic energy consumption, for the 7-points stencil matrix with a total of $408^3$ DOFs under weak scalability. Columns report GPU, CPU, and total dynamic energy consumption expressed as a percentage of the static energy. }
\label{tab:solver_dynamicVSstatic_noprec7p}
\begin{tabular}{llrrr}
\toprule
 &  & GPU \% & CPU \% & total \% \\
\#GPUs & library &  &  &  \\
\midrule
\multirow[t]{3}{*}{1} & AMGX & 217.33 & 12.40 & 105.53 \\
 & BCMGX & 81.26 & 12.44 & 59.17 \\
 & Ginkgo & 111.99 & 11.07 & 96.67 \\
\cline{1-5}
\multirow[t]{3}{*}{2} & AMGX & 186.69 & 23.56 & 105.86 \\
 & BCMGX & 69.32 & 22.96 & 55.77 \\
 & Ginkgo & 84.84 & 23.44 & 77.59 \\
\cline{1-5}
\multirow[t]{3}{*}{4} & AMGX & 172.72 & 19.62 & 99.63 \\
 & BCMGX & 71.69 & 16.71 & 55.60 \\
 & Ginkgo & 114.80 & 14.98 & 99.56 \\
\cline{1-5}
\multirow[t]{3}{*}{8} & AMGX & 170.74 & 19.83 & 99.01 \\
 & BCMGX & 63.09 & 16.93 & 50.67 \\
 & Ginkgo & 94.92 & 21.31 & 85.37 \\
\cline{1-5}
\multirow[t]{3}{*}{16} & AMGX & 166.05 & 16.61 & 95.87 \\
 & BCMGX & 62.91 & 18.87 & 51.18 \\
 & Ginkgo & 100.67 & 17.26 & 89.14 \\
\cline{1-5}
\multirow[t]{3}{*}{32} & AMGX & 169.68 & 17.20 & 96.55 \\
 & BCMGX & 71.43 & 17.84 & 55.81 \\
 & Ginkgo & 104.61 & 19.48 & 92.11 \\
\cline{1-5}
\multirow[t]{3}{*}{64} & AMGX & 161.78 & 18.45 & 94.16 \\
 & BCMGX & 63.73 & 18.15 & 51.49 \\
 & Ginkgo & 102.88 & 18.91 & 90.60 \\
\cline{1-5}
\bottomrule
\end{tabular}
\end{table}

\begin{table}
\centering
\caption{Static vs. dynamic energy consumption, for the 27-points stencil matrix with a total of $265^3$ DOFs under weak scalability. Columns report GPU, CPU, and total dynamic energy consumption expressed as a percentage of the static energy. }
\label{tab:solver_dynamicVSstatic_noprec27p}
\begin{tabular}{llrrr}
\toprule
 &  & GPU \% & CPU \% & total \% \\
\#GPUs & library &  &  &  \\
\midrule
\multirow[t]{2}{*}{1} & BCMGX & 47.04 & 9.60 & 40.27 \\
 & Ginkgo & 77.70 & 14.69 & 71.42 \\
\cline{1-5}
\multirow[t]{2}{*}{2} & BCMGX & 46.94 & 17.01 & 41.38 \\
 & Ginkgo & 84.94 & 23.38 & 78.34 \\
\cline{1-5}
\multirow[t]{2}{*}{4} & BCMGX & 40.38 & 19.42 & 36.82 \\
 & Ginkgo & 82.22 & 11.40 & 74.74 \\
\cline{1-5}
\multirow[t]{2}{*}{8} & BCMGX & 47.12 & 19.67 & 42.05 \\
 & Ginkgo & 89.56 & 19.81 & 81.61 \\
\cline{1-5}
\multirow[t]{2}{*}{16} & BCMGX & 42.46 & 19.65 & 38.53 \\
 & Ginkgo & 83.12 & 20.80 & 76.59 \\
\cline{1-5}
\multirow[t]{2}{*}{32} & BCMGX & 40.76 & 19.43 & 37.15 \\
 & Ginkgo & 79.44 & 17.69 & 73.35 \\
\cline{1-5}
\multirow[t]{2}{*}{64} & BCMGX & 41.71 & 17.89 & 37.77 \\
 & Ginkgo & 78.55 & 18.49 & 72.59 \\
\cline{1-5}
\bottomrule
\end{tabular}
\end{table}

\begin{figure}[h]
\centering
     \begin{subfigure}[t]{0.48\textwidth}
         \centering
         \includegraphics[width=\textwidth]{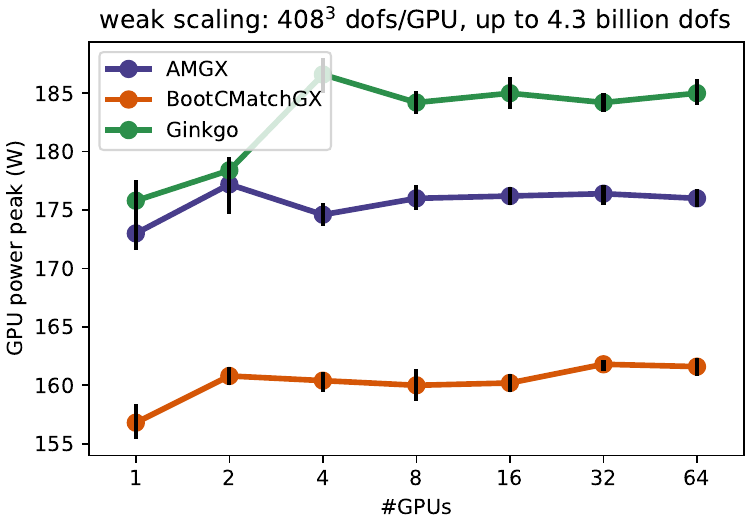}
         \caption{7-points stencil matrix with $408^3$ DOFs per GPU under weak scalability.}
         \label{fig:cg_powerWeak7}
     \end{subfigure}
     \hfill
    \begin{subfigure}[t]{0.48\textwidth}
         \centering
         \includegraphics[width=\textwidth]{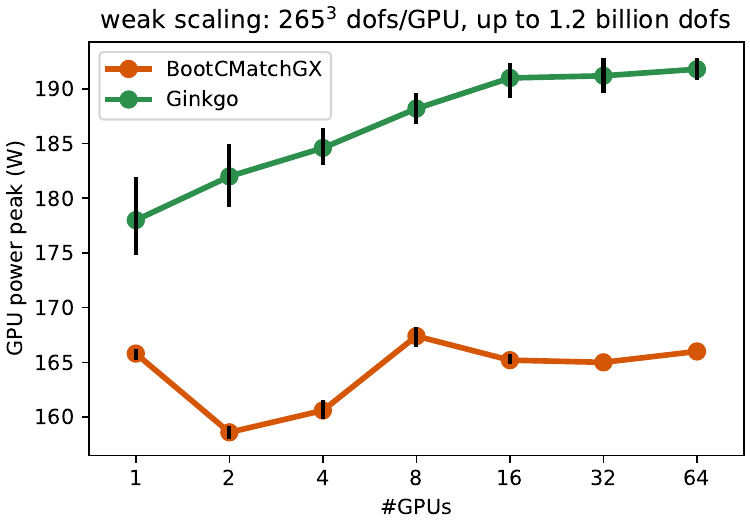}
         \caption{27-points stencil matrix with $265^3$ DOFs per GPU under weak scalability.}
         \label{fig:cg_powerWeak27}
     \end{subfigure}
     \hfill
     \begin{subfigure}[t]{0.48\textwidth}
         \centering
         \includegraphics[width=\textwidth]{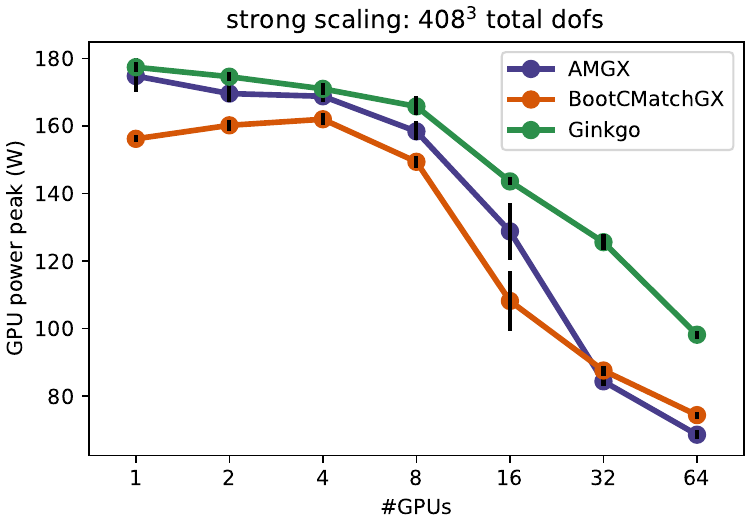}
         \caption{7-points stencil matrix with a total of $408^3$ DOFs under strong scalability.}
         \label{fig:cg_powerStrong7}
     \end{subfigure}
     \hfill
     \begin{subfigure}[t]{0.48\textwidth}
         \centering
         \includegraphics[width=\textwidth]{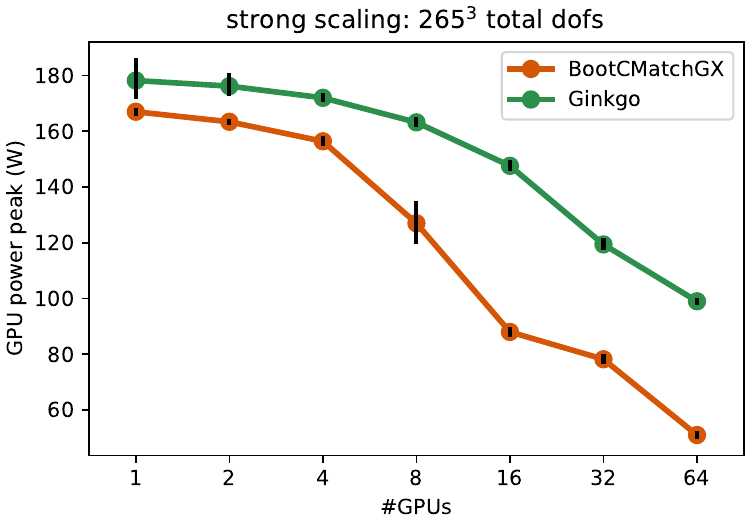}
         \caption{27-points stencil matrix with a total of $265^3$ DOFs under strong scalability.}
         \label{fig:cg_powerStrong27}
     \end{subfigure}
\caption{GPU power peak of the CG computation under weak and strong scalability scenarios.} 
\label{fig:cg_power}
\end{figure}
Finally, Figure~\ref{fig:cg_power} shows the GPU power peaks recorded during the CG computation. \texttt{BootCMatchGX} consistently exhibits lower peaks than both Ginkgo and NVIDIA AmgX. Under strong scalability, these peaks decrease as the number of GPUs increases, reflecting the reduced workload per GPU. The higher power peaks observed for Ginkgo, in both weak and strong scalability, help explain its higher energy consumption even when, in the strong scaling scenario, execution times are comparable. Overall, this suggests that \texttt{BootCMatchGX} manages GPU resources more efficiently, maintaining a steady workload and avoiding sudden spikes in power demand.

\subsubsection{Preconditioned Conjugate Gradient solver}
\label{pcg}

We evaluate the PCG implementations of NVIDIA AmgX and \texttt{BootCMatchGX} under strong and weak scalability conditions. For fairness, AmgX is configured with the matching-based aggregation preconditioner, using aggregates of size 8, as in \texttt{BootCMatchGX}. Both libraries use default settings for the AMG hierarchy (levels and coarsest size) and apply the same smoother (4 $\ell_1$-Jacobi iterations) in the V-cycle.
For the single-GPU case, the problem size is set to $370^3$  DOFs (7-point stencil). 
All performance measurements reported in the figures represent averages over five independent runs and include 95\% confidence intervals, shown as black vertical lines.  For these tests, iterations stop at a relative residual tolerance of  $10^{-6}$, reflecting realistic simulation settings.

\begin{figure}[ht]
\centering
     \begin{subfigure}[t]{0.48\textwidth}
         \centering
         \includegraphics[width=\textwidth]{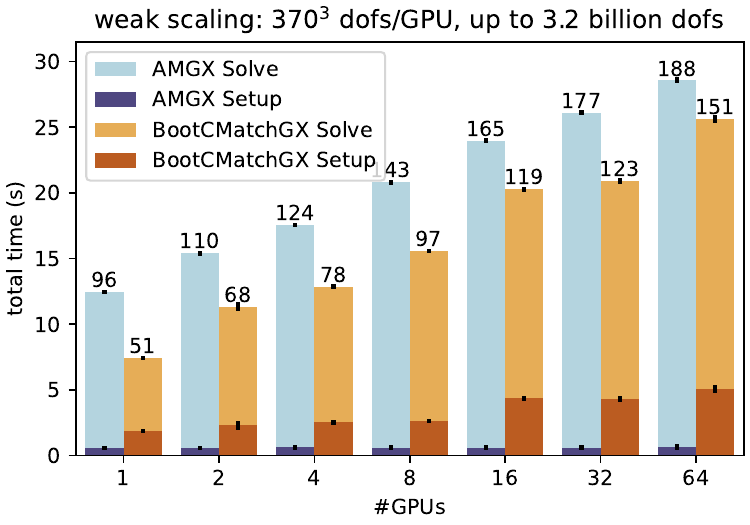}
         \caption{7-points stencil matrix with $370^3$ DOFs pr GPU under weak scalability.}
     \end{subfigure}
     \hfill
     \begin{subfigure}[t]{0.48\textwidth}
         \centering
         \includegraphics[width=\textwidth]{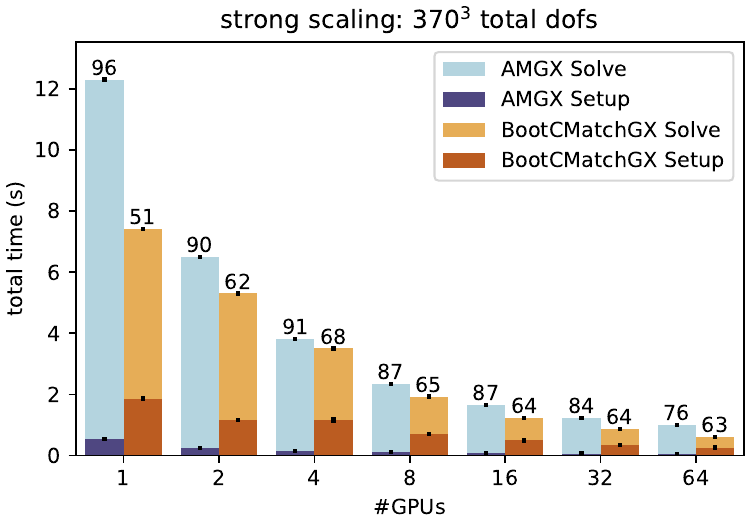}
         \caption{7-points stencil matrix with a total of $370^3$ DOFs under strong scalability.}
     \end{subfigure}
\caption{Execution times breakdown of the PCG method of solve and setup times under weak and strong scalability scenarios. The number on top of each bar denotes the number of iterations carried out by solvers.} 
\label{fig:pcg_time}
\end{figure}
\begin{figure}[ht]
\centering
     \begin{subfigure}[t]{0.48\textwidth}
         \centering
         \includegraphics[width=\textwidth]{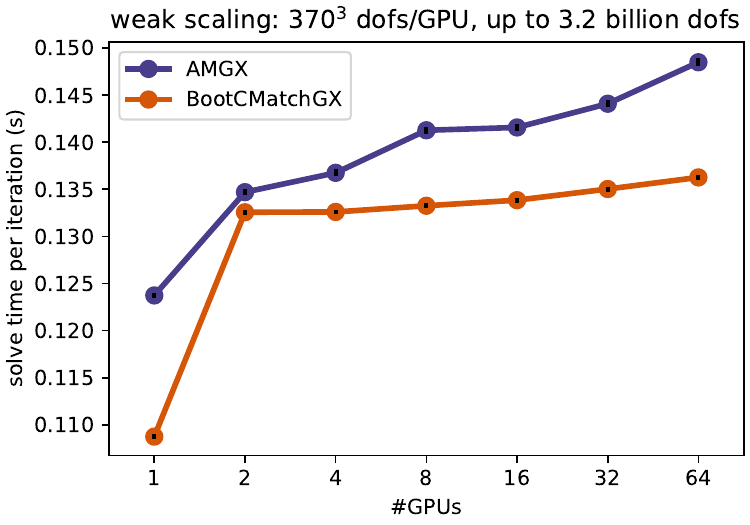}
         \caption{7-points stencil matrix with $370^3$ DOFs per GPU under weak scalability.}
     \end{subfigure}
     \hfill
     \begin{subfigure}[t]{0.48\textwidth}
         \centering
         \includegraphics[width=\textwidth]{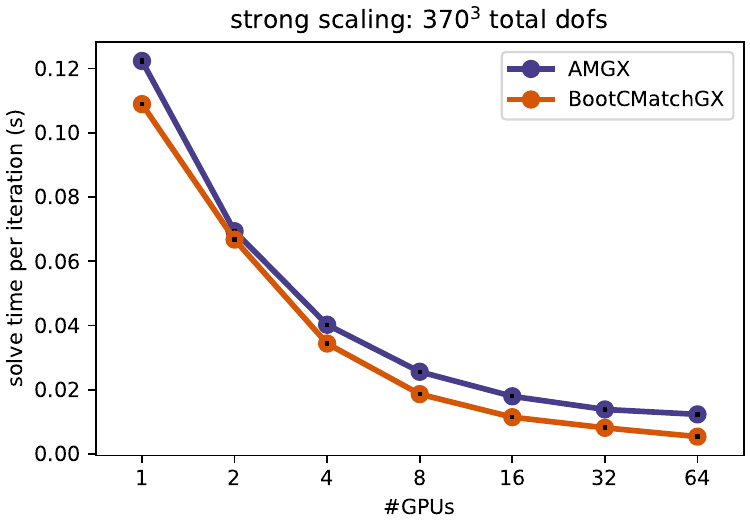}
         \caption{7-points stencil matrix with a total of $370^3$ DOFs under strong scalability.}
     \end{subfigure}
\caption{Solve time per iteration of the PCG method under weak and strong scalability scenarios.} 
\label{fig:pcg_time_iteration}
\end{figure}
Figure~\ref{fig:pcg_time} breaks down the PCG execution times in setup and solve phases shown as colored segments in each bar. The results indicate that \texttt{BootCMatchGX} consistently outperforms NVIDIA AmgX by achieving lower runtimes. This improvement stems from the \texttt{BootCMatchGX} preconditioner, which reduces the number of solver iterations and thus provides a clear reduction in the solution time. Moreover, as shown in Figure~\ref{fig:pcg_time_iteration}, the per-iteration solve time of \texttt{BootCMatchGX} also remains consistently lower than that of NVIDIA AmgX, thereby confirming the performance advantage of \texttt{BootCMatchGX} over NVIDIA AmgX.

\begin{figure}[h]
\centering
     \begin{subfigure}[t]{0.48\textwidth}
         \centering
         \includegraphics[width=\textwidth]{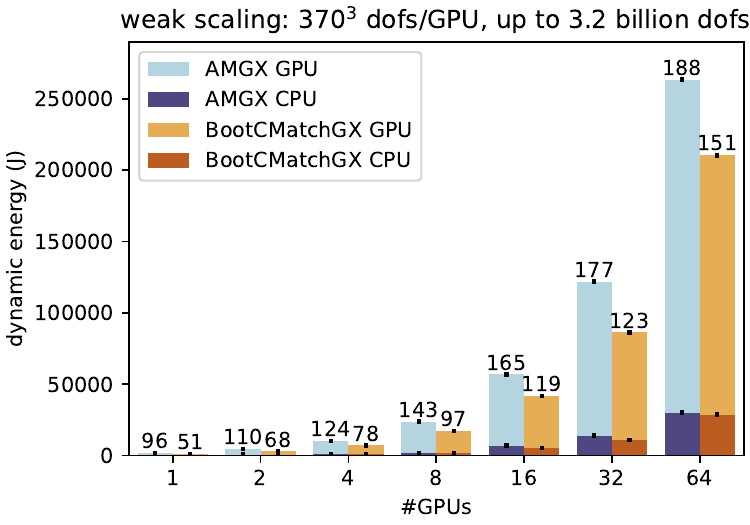}
         \caption{7-points stencil matrix with $370^3$ DOFs per GPU under weak scalability.}
         \label{fig:pcg_energyWeak7}
     \end{subfigure}
     \hfill
     \begin{subfigure}[t]{0.48\textwidth}
         \centering
         \includegraphics[width=\textwidth]{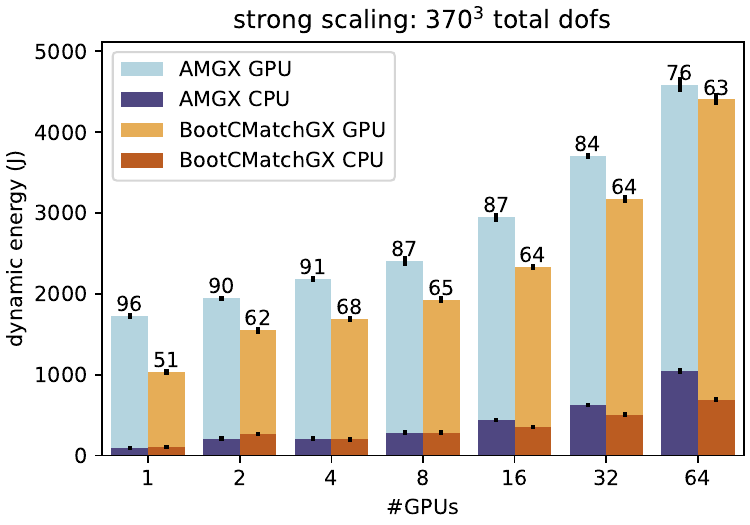}
         \caption{7-points stencil matrix with a total of $370^3$ DOFs under strong scalability.}
         \label{fig:pcg_energyStrong7}
     \end{subfigure}
\caption{Dynamic energy consumption breakdown of the PCG computation on GPU and CPU under weak and strong scalability scenarios. 
} 
\label{fig:pcg_energy}
\end{figure}
Figure~\ref{fig:pcg_energy} breaks down the dynamic energy consumption for the PCG computation, with GPU and CPU contributions shown as colored segments in each bar. The results confirm the trends observed in Figure~\ref{fig:pcg_time}: \texttt{BootCMatchGX} is consistently more energy‑efficient than NVIDIA AmgX. In every scenario, its dynamic energy consumption is lower than that of NVIDIA AmgX. As noted above, the lower iteration count of \texttt{BootCMatchGX} and per-iteration solve time yield an energy‑saving advantage over NVIDIA AmgX.

\begin{figure}[h]
\centering
     \begin{subfigure}[t]{0.48\textwidth}
         \centering
         \includegraphics[width=\textwidth]{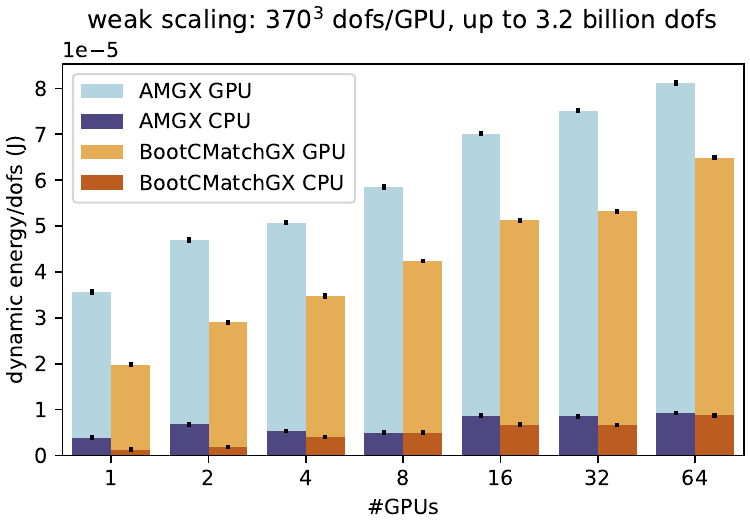}
         \caption{7-points stencil matrix with $370^3$ DOFs per GPU under weak scalability.}
         \label{fig:pcg_energyDofsWeak7}
     \end{subfigure}
     \hfill
     \begin{subfigure}[t]{0.48\textwidth}
         \centering
         \includegraphics[width=\textwidth]{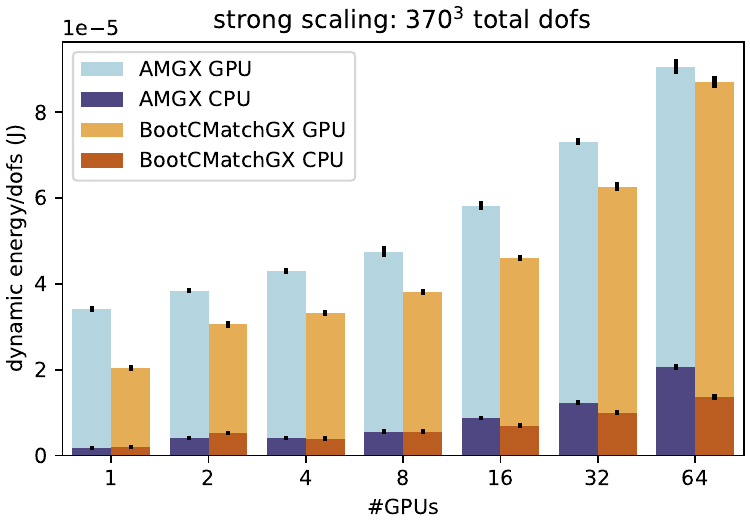}
         \caption{7-points stencil matrix with a total of $370^3$ DOFs under strong scalability.}
         \label{fig:pcg_energyDofsStrong7}
     \end{subfigure}
\caption{Dynamic energy consumption per DOF breakdown of the PCG computation on GPU and CPU under weak and strong scalability scenarios.} 
\label{fig:pcg_energyDofs}
\end{figure}
Figure~\ref{fig:pcg_energyDofs} presents the breakdown of dynamic energy consumption per DOF for the PCG computation, with GPU and CPU contributions shown as colored segments in each bar. Unlike the un-preconditioned CG case, and as expected given the additional cost of applying the preconditioner, we observe an increase in per-DOF energy consumption under both weak and strong scaling. Although the growth is gradual, it becomes noticeable as the number of tasks increases proportionally to the global problem size. The effect is more pronounced in the strong scaling scenario, where the reduced workload per GPU leads to lower computational efficiency. In all cases, however, \texttt{BootCMatchGX} remains consistently more energy-efficient than NVIDIA AmgX under both weak and strong scalability.

\begin{figure}[h]
\centering
     \begin{subfigure}[t]{0.48\textwidth}
         \centering
         \includegraphics[width=\textwidth]{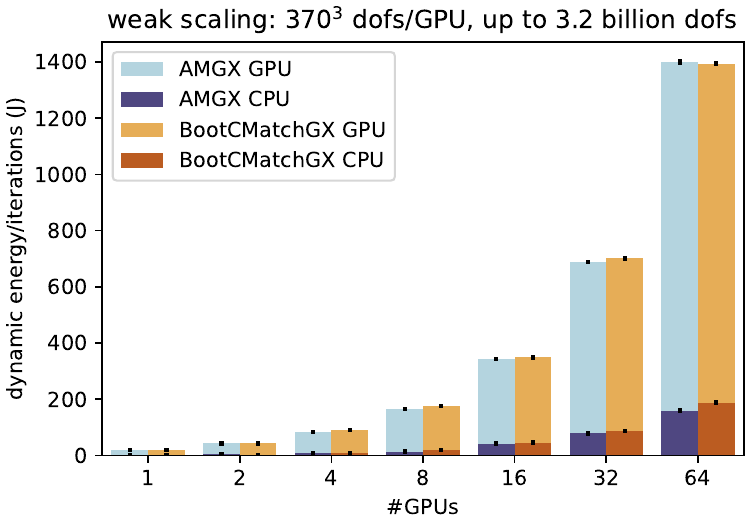}
         \caption{7-points stencil matrix with $370^3$ DOFs per GPU under weak scalability.}
         \label{fig:pcg_energyIterWeak7}
     \end{subfigure}
     \hfill
     \begin{subfigure}[t]{0.48\textwidth}
         \centering
         \includegraphics[width=\textwidth]{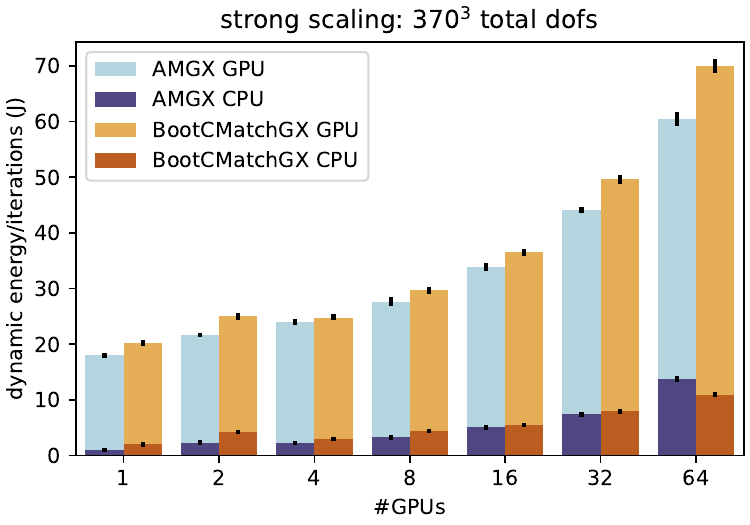}
         \caption{7-points stencil matrix with a total of $370^3$ DOFs under strong scalability.}
         \label{fig:pcg_energyIterStrong7}
     \end{subfigure}
\caption{Dynamic energy consumption per iteration breakdown of the PCG computation on GPU and CPU under weak and strong scalability scenarios.} 
\label{fig:pcg_energyIter}
\end{figure}
Figure~\ref{fig:pcg_energyIter} presents the breakdown of dynamic energy consumption per iteration for the PCG computation, with GPU and CPU contributions shown as colored segments in each bar. Under weak scaling, the per-iteration energy increases moderately as the number of processes grows. This trend is consistent with the growing cost of communication/synchronization and the parallel overheads of applying the preconditioner at larger scales. Notably, the weak-scaling trends of \texttt{BootCMatchGX} and NVIDIA AmgX are essentially indistinguishable, indicating that the impact of the preconditioner application is equivalent for both implementations. In the strong-scaling scenario, as the workload per GPU diminishes, NVIDIA AmgX exhibits a slight advantage in energy efficiency over \texttt{BootCMatchGX} at higher process counts.

\begin{figure}[h]
\centering
     \begin{subfigure}[t]{0.48\textwidth}
         \centering
         \includegraphics[width=\textwidth]{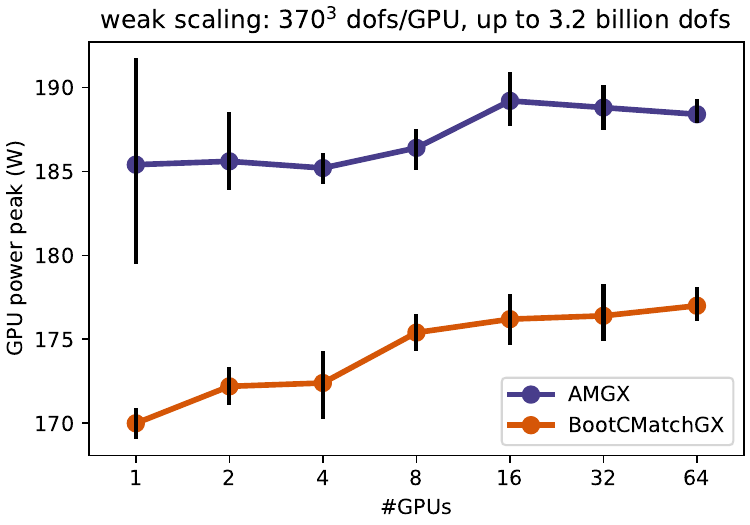}
         \caption{7-points stencil matrix with $370^3$ DOFs per GPU under weak scalability.}
         \label{fig:pcg_powerWeak7}
     \end{subfigure}
     \hfill
     \begin{subfigure}[t]{0.48\textwidth}
         \centering
         \includegraphics[width=\textwidth]{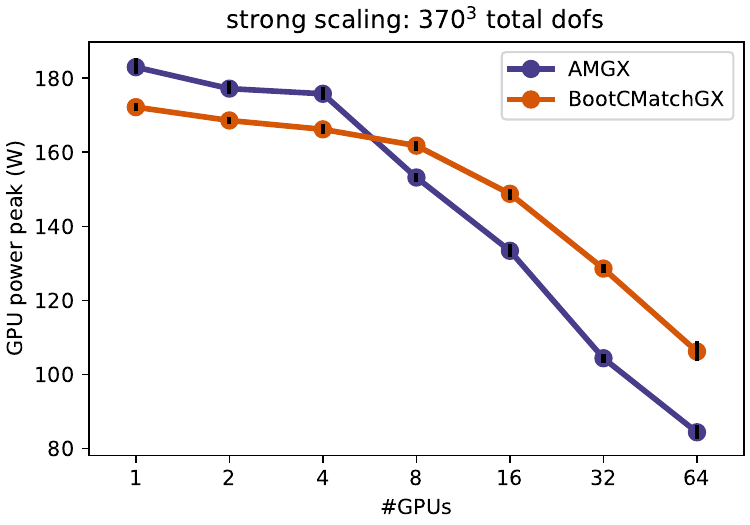}
         \caption{7-points stencil matrix with a total of $370^3$ DOFs under strong scalability.}
         \label{fig:pcg_powerStrong7}
     \end{subfigure}
\caption{GPU power peak of the PCG computation under weak and strong scalability scenarios.} 
\label{fig:pcg_power}
\end{figure}
Figure~\ref{fig:pcg_power} shows the GPU power peaks recorded during the PCG computation. Under weak scalability, \texttt{BootCMatchGX} consistently exhibits lower peaks than NVIDIA AmgX, suggesting that it manages GPU resources more efficiently by maintaining a steady workload and avoiding sudden spikes in power demand. Under strong scalability, power peaks decrease as the number of GPUs increases, reflecting the reduced workload per GPU. However, starting from 8 GPUs, NVIDIA AmgX exhibits lower peaks than \texttt{BootCMatchGX}. This behavior is reflected in the higher per-iteration energy consumption of \texttt{BootCMatchGX}, as shown in Figure~\ref{fig:pcg_energyIter}.

Finally, Table~\ref{tab:solver_dynamicVSstatic_prec7p} compares static and dynamic energy consumption, expressing the latter as a percentage of the former. AmgX exhibits higher dynamic energy percentages than \texttt{BootCMatchGX}, together with greater dynamic energy consumption and longer runtimes, reflecting less efficient GPU utilization and resulting in higher total energy consumption. 

\begin{table}
\centering
\caption{Static vs. dynamic energy consumption, for the 7-points stencil matrix with a total of $370^3$ DOFs under weak scalability. Columns report GPU, CPU, and total dynamic energy consumption expressed as a percentage of the static energy. }
\label{tab:solver_dynamicVSstatic_prec7p}
\begin{tabular}{llrrr}
\toprule
 &  & GPU \% & CPU \% & total \% \\
\#GPUs & library &  &  &  \\
\midrule
\multirow[t]{2}{*}{1} & AMGX & 262.39 & 18.40 & 106.57 \\
 & BootCMatchGX & 155.01 & 10.11 & 80.18 \\
\cline{1-5}
\multirow[t]{2}{*}{2} & AMGX & 294.45 & 25.88 & 117.24 \\
 & BootCMatchGX & 172.49 & 9.58 & 82.48 \\
\cline{1-5}
\multirow[t]{2}{*}{4} & AMGX & 301.32 & 17.68 & 112.13 \\
 & BootCMatchGX & 226.17 & 18.02 & 97.38 \\
\cline{1-5}
\multirow[t]{2}{*}{8} & AMGX & 309.09 & 14.02 & 110.24 \\
 & BootCMatchGX & 237.75 & 18.66 & 99.85 \\
\cline{1-5}
\multirow[t]{2}{*}{16} & AMGX & 329.15 & 21.06 & 117.17 \\
 & BootCMatchGX & 218.67 & 19.35 & 92.89 \\
\cline{1-5}
\multirow[t]{2}{*}{32} & AMGX & 327.52 & 19.15 & 115.29 \\
 & BootCMatchGX & 238.84 & 18.56 & 96.01 \\
\cline{1-5}
\multirow[t]{2}{*}{64} & AMGX & 323.88 & 18.86 & 113.92 \\
 & BootCMatchGX & 241.09 & 19.93 & 96.36 \\
\cline{1-5}
\bottomrule
\end{tabular}
\end{table}

Overall, our analysis shows that \texttt{BootCMatchGX} outperforms NVIDIA AmgX in both execution time and energy consumption, thereby reducing the overall energy footprint. The only exception occurs in the strong scaling scenario, where NVIDIA AmgX demonstrates a slight advantage in per-iteration dynamic energy consumption as the number of GPUs increases.

\subsection{Results on Matrices from SuiteSparse}
\label{suitesparse}

For space reasons, we restrict the discussion to the main results obtained on matrices from the SuiteSparse Collection, summarizing execution times and dynamic energy consumption (GPU, CPU, and total), as well as peak GPU power, for both the SpMV operation and the CG solver.
We do not report results obtained using the AMG preconditioners available in the libraries, since AMG methods are not the most appropriate preconditioners for general SPD matrices that do not arise from diffusion-dominated problems, such as those stemming from structural mechanics or optimization applications. Moreover, a detailed analysis of the convergence behavior of AMG preconditioners on general matrices is beyond the scope of this paper.

\begin{table*}
\centering
\caption{Results with the SpMV Operation on the matrices of Table~\ref{tab:suitesparse_matrices}.\label{tab:spdmatspmv}}
\begin{tabular}{lllm{15mm}m{14mm}m{14mm}m{14mm}m{14mm}}
\toprule
 &  &  & time (s) & GPU\newline dynamic energy (J) & CPU\newline dynamic energy (J) & Total\newline dynamic energy (J) & GPU power peak (W) \\
\# GPUs & matrix & library &  &  &  &  &  \\
\midrule
\multirow[t]{10}{*}{1} & \multirow[t]{2}{*}{G3\_circuit} & BCMGX & 0.0002 & 0.1637 & 0.0028 & 0.1665 & 82 \\
 &  & Ginkgo & 0.0004 & 0.2011 & 0.0069 & 0.2080 & 64 \\
\cline{2-8}
 & \multirow[t]{2}{*}{af\_shell8} & BCMGX & 0.0004 & 0.1907 & 0.0029 & 0.1936 & 100 \\
 &  & Ginkgo & 0.0006 & 0.5047 & 0.0078 & 0.5125 & 100 \\
\cline{2-8}
 & \multirow[t]{2}{*}{boneS10} & BCMGX & 0.0012 & 0.6754 & 0.0115 & 0.6869 & 105 \\
 &  & Ginkgo & 0.0012 & 1.1370 & 0.0170 & 1.1540 & 197 \\
\cline{2-8}
 & \multirow[t]{2}{*}{ecology2} & BCMGX & 0.0001 & 0.0557 & 0.0023 & 0.0579 & 63 \\
 &  & Ginkgo & 0.0003 & 0.1363 & 0.0042 & 0.1405 & 64 \\
\cline{2-8}
 & \multirow[t]{2}{*}{parabolic\_fem} & BCMGX & 0.0001 & 0.1333 & 0.0021 & 0.1354 & 40 \\
 &  & Ginkgo & 0.0002 & 0.0637 & 0.0024 & 0.0662 & 59 \\
\cline{1-8} \cline{2-8}
\multirow[t]{10}{*}{2} & \multirow[t]{2}{*}{G3\_circuit} & BCMGX & 0.0002 & 0.2180 & 0.0096 & 0.2275 & 61 \\
 &  & Ginkgo & 0.0013 & 0.4741 & 0.0620 & 0.5361 & 46 \\
\cline{2-8}
 & \multirow[t]{2}{*}{af\_shell8} & BCMGX & 0.0002 & 0.2896 & 0.0070 & 0.2965 & 76 \\
 &  & Ginkgo & 0.0005 & 0.8035 & 0.0175 & 0.8210 & 82 \\
\cline{2-8}
 & \multirow[t]{2}{*}{boneS10} & BCMGX & 0.0007 & 0.7326 & 0.0236 & 0.7562 & 102 \\
 &  & Ginkgo & 0.0009 & 2.0074 & 0.0285 & 2.0359 & 179 \\
\cline{2-8}
 & \multirow[t]{2}{*}{ecology2} & BCMGX & 0.0001 & 0.1178 & 0.0040 & 0.1218 & 48 \\
 &  & Ginkgo & 0.0002 & 0.2628 & 0.0104 & 0.2733 & 64 \\
\cline{2-8}
 & \multirow[t]{2}{*}{parabolic\_fem} & BCMGX & 0.0004 & 0.1641 & 0.0183 & 0.1824 & 42 \\
 &  & Ginkgo & 0.0018 & 0.2526 & 0.1061 & 0.3587 & 41 \\
\cline{1-8} \cline{2-8}
\multirow[t]{10}{*}{4} & \multirow[t]{2}{*}{G3\_circuit} & BCMGX & 0.0004 & 0.3660 & 0.0311 & 0.3971 & 43 \\
 &  & Ginkgo & 0.0015 & 1.1603 & 0.1143 & 1.2746 & 52 \\
\cline{2-8}
 & \multirow[t]{2}{*}{af\_shell8} & BCMGX & 0.0001 & 0.6245 & 0.0095 & 0.6340 & 67 \\
 &  & Ginkgo & 0.0003 & 1.6336 & 0.0218 & 1.6554 & 68 \\
\cline{2-8}
 & \multirow[t]{2}{*}{boneS10} & BCMGX & 0.0003 & 1.1179 & 0.0265 & 1.1444 & 93 \\
 &  & Ginkgo & 0.0007 & 3.7145 & 0.0479 & 3.7624 & 146 \\
\cline{2-8}
 & \multirow[t]{2}{*}{ecology2} & BCMGX & 0.0001 & 0.3321 & 0.0066 & 0.3387 & 45 \\
 &  & Ginkgo & 0.0002 & 0.7931 & 0.0113 & 0.8044 & 74 \\
\cline{2-8}
 & \multirow[t]{2}{*}{parabolic\_fem} & BCMGX & 0.0007 & 0.4641 & 0.0695 & 0.5336 & 40 \\
 &  & Ginkgo & 0.0028 & 0.9583 & 0.2921 & 1.2504 & 58 \\
\cline{1-8} \cline{2-8}
\bottomrule
\end{tabular}
\end{table*}

Table~\ref{tab:spdmatspmv} summarizes the SpMV performance results for
BootCMatchGX (BCMGX) and Ginkgo. As expected, the scalability trend strongly depends on
the sparsity pattern and average number of nonzeros per row. Matrices with a
more regular structure and higher average connectivity, such as
\texttt{boneS10} and \texttt{af\_shell8}, exhibit better scalability, whereas
matrices characterized by irregular sparsity patterns, such as \texttt{G3$\_$circuit}, show an increase
in execution time for increasing number of GPUs due to increased communication and load imbalance effects. The \texttt{parabolic$\_$fem} matrix exhibits a regular sparsity pattern typical of FEM
discretizations, however its SpMV performance degrades as the number of GPUs increases, due to
the presence of non-zeros far from the diagonal, which leads to increased
inter-process communication that negatively impacts scalability. The \texttt{ecology2} matrix exhibits a moderately irregular sparsity pattern with
most non-zeros located close to the diagonal, resulting in limited but
non-negligible inter-process communication in multi-GPU SpMV.
The energy analysis closely follows the observed runtime behavior. 
BootCMatchGX exhibits low energy consumption across all configurations except one, the single-GPU \texttt{parabolic\_fem} case. Its performance is consistent with shorter runtimes and more effective utilization of GPU resources, as indicated by lower power peaks and average power. The only exception is characterized by higher average power, which leads to greater energy consumption despite a shorter runtime compared to Ginkgo.
Regarding GPU power peaks, the results indicate a moderate decrease as the number
of GPUs grows, consistently with the reduced workload per device. Across all
test cases, BootCMatchGX exhibits lower or comparable power peaks than Ginkgo,
suggesting a more stable utilization of GPU resources and a reduced incidence of
short-lived power spikes.
Overall, the comparison highlights that BootCMatchGX generally outperforms
Ginkgo for the SpMV operation also on SuiteSparse matrices, achieving lower execution
times, reduced dynamic energy consumption, and more favorable power profiles.
These results confirm that the optimizations adopted in BootCMatchGX translate
into tangible benefits not only in performance, but also in energy efficiency
across a wide range of sparse matrix characteristics.

\begin{table*}
\centering
\caption{Results with the CG Solver on the matrices of Table~\ref{tab:suitesparse_matrices}.\label{tab:spdmatcg}}
\begin{tabular}{lllm{15mm}m{14mm}m{14mm}m{14mm}m{14mm}}
\toprule
 &  &  & runtime (s) & GPU\newline dynamic energy (J) & CPU\newline dynamic energy (J) & dynamic energy (J) & GPU power peak (W) \\
\# GPUs & matrix & library &  &  &  &  &  \\
\midrule
\multirow[t]{15}{*}{1} & \multirow[t]{3}{*}{G3\_circuit} & AMGX & 0.0686 & 18.7061 & 1.1757 & 19.8818 & 88 \\
 &  & BCMGX & 0.0588 & 8.4993 & 0.8437 & 9.3430 & 82 \\
 &  & Ginkgo & 0.0804 & 61.5499 & 0.7746 & 62.3245 & 143 \\
\cline{2-8}
 & \multirow[t]{3}{*}{af\_shell8} & AMGX & 0.0660 & 30.6657 & 1.0121 & 31.6779 & 91 \\
 &  & BCMGX & 0.0643 & 12.3762 & 1.0650 & 13.4412 & 83 \\
 &  & Ginkgo & 0.0701 & 75.3592 & 1.2008 & 76.5601 & 163 \\
\cline{2-8}
 & \multirow[t]{3}{*}{boneS10} & AMGX & 0.1199 & 77.2129 & 1.6811 & 78.8940 & 100 \\
 &  & BCMGX & 0.1478 & 61.1318 & 2.1697 & 63.3015 & 98 \\
 &  & Ginkgo & 0.1527 & 166.6234 & 0.8179 & 167.4413 & 175 \\
\cline{2-8}
 & \multirow[t]{3}{*}{ecology2} & AMGX & 0.0503 & 12.0110 & 0.5640 & 12.5750 & 70 \\
 &  & BCMGX & 0.0434 & 2.6365 & 0.6477 & 3.2843 & 64 \\
 &  & Ginkgo & 0.0612 & 53.7802 & 0.6177 & 54.3979 & 122 \\
\cline{2-8}
 & \multirow[t]{3}{*}{parabolic\_fem} & AMGX & 0.0373 & 13.0647 & 0.4709 & 13.5357 & 63 \\
 &  & BCMGX & 0.0330 & 3.6301 & 0.5339 & 4.1640 & 63 \\
 &  & Ginkgo & 0.0490 & 24.3215 & 0.7923 & 25.1138 & 85 \\
\cline{1-8} \cline{2-8}
\multirow[t]{15}{*}{2} & \multirow[t]{3}{*}{G3\_circuit} & AMGX & 0.0674 & 28.3306 & 2.6856 & 31.0162 & 58 \\
 &  & BCMGX & 0.0559 & 12.4181 & 2.7095 & 15.1275 & 63 \\
 &  & Ginkgo & 0.1671 & 60.6846 & 6.1768 & 66.8614 & 52 \\
\cline{2-8}
 & \multirow[t]{3}{*}{af\_shell8} & AMGX & 0.0501 & 40.8106 & 2.6652 & 43.4758 & 68 \\
 &  & BCMGX & 0.0470 & 17.9323 & 2.2262 & 20.1585 & 76 \\
 &  & Ginkgo & 0.0663 & 135.9255 & 1.8674 & 137.7929 & 121 \\
\cline{2-8}
 & \multirow[t]{3}{*}{boneS10} & AMGX & 0.0755 & 141.1463 & 3.7273 & 144.8736 & 91 \\
 &  & BCMGX & 0.0943 & 52.7114 & 3.6241 & 56.3355 & 90 \\
 &  & Ginkgo & 0.1132 & 260.4663 & 4.0540 & 264.5203 & 158 \\
\cline{2-8}
 & \multirow[t]{3}{*}{ecology2} & AMGX & 0.0424 & 11.2354 & 2.0118 & 13.2473 & 56 \\
 &  & BCMGX & 0.0341 & 5.4043 & 1.2162 & 6.6206 & 44 \\
 &  & Ginkgo & 0.0533 & 57.3245 & 1.4502 & 58.7747 & 74 \\
\cline{2-8}
 & \multirow[t]{3}{*}{parabolic\_fem} & AMGX & 0.0736 & 14.4014 & 3.6070 & 18.0084 & 44 \\
 &  & BCMGX & 0.0670 & 7.2354 & 2.7290 & 9.9644 & 42 \\
 &  & Ginkgo & 0.2285 & 29.8936 & 11.4415 & 41.3350 & 44 \\
\cline{1-8} \cline{2-8}
\multirow[t]{15}{*}{4} & \multirow[t]{3}{*}{G3\_circuit} & AMGX & 0.0707 & 53.4715 & 5.0398 & 58.5113 & 45 \\
 &  & BCMGX & 0.0645 & 19.6857 & 4.7947 & 24.4804 & 46 \\
 &  & Ginkgo & 0.2034 & 83.7765 & 17.5171 & 101.2936 & 59 \\
\cline{2-8}
 & \multirow[t]{3}{*}{af\_shell8} & AMGX & 0.0405 & 92.8779 & 3.3988 & 96.2767 & 57 \\
 &  & BCMGX & 0.0343 & 35.9482 & 2.1871 & 38.1353 & 59 \\
 &  & Ginkgo & 0.0608 & 164.4908 & 4.0691 & 168.5599 & 83 \\
\cline{2-8}
 & \multirow[t]{3}{*}{boneS10} & AMGX & 0.0543 & 194.7348 & 5.4353 & 200.1701 & 76 \\
 &  & BCMGX & 0.0611 & 75.3093 & 4.2240 & 79.5334 & 78 \\
 &  & Ginkgo & 0.0832 & 454.3023 & 4.9802 & 459.2825 & 134 \\
\cline{2-8}
 & \multirow[t]{3}{*}{ecology2} & AMGX & 0.0375 & 30.6689 & 2.4062 & 33.0751 & 44 \\
 &  & BCMGX & 0.0290 & 10.6009 & 1.6763 & 12.2772 & 43 \\
 &  & Ginkgo & 0.0510 & 75.3333 & 2.7597 & 78.0929 & 69 \\
\cline{2-8}
 & \multirow[t]{3}{*}{parabolic\_fem} & AMGX & 0.0946 & 28.1975 & 9.7096 & 37.9071 & 39 \\
 &  & BCMGX & 0.0894 & 19.5766 & 7.6658 & 27.2424 & 40 \\
 &  & Ginkgo & 0.3167 & 66.5021 & 25.2880 & 91.7900 & 52 \\
\cline{1-8} \cline{2-8}
\bottomrule
\end{tabular}
\end{table*}

The results reported in Table~\ref{tab:spdmatcg} show trends that are fully
consistent with those observed for the SpMV operation in Table~\ref{tab:spdmatspmv}.
This behavior is expected, since SpMV represents the dominant computational
kernel within each CG iteration, accounting for the largest fraction of both
execution time and dynamic energy consumption.
As the number of GPUs increases, the CG execution time follows the same scaling
patterns observed for SpMV across the different matrices. In particular,
matrices exhibiting favorable SpMV scalability, such as \texttt{boneS10} and \texttt{af$\_$shell8},
also show improved CG performance, whereas matrices characterized by irregular
sparsity patterns or large off-diagonal contributions, such as \texttt{G3$\_$circuit} and
\texttt{parabolic$\_$fem}, exhibit a degradation in CG scalability due to increased
communication overheads. The \texttt{ecology2} matrix again represents an intermediate
case, with moderate scalability limitations reflecting its sparsity structure.
The energy consumption trends closely mirror the runtime behavior. Since the
energy-to-solution of CG is largely dictated by the cumulative cost of repeated
SpMV operations, reductions (or increases) in execution time directly translate
into corresponding changes in dynamic energy consumption. Similarly, GPU power
peaks show the same qualitative behavior observed for SpMV, decreasing with the
number of GPUs as the per-device workload is reduced. Overall, the CG results confirm that improvements achieved at the SpMV level
carry over directly to the full solver. In this respect, BootCMatchGX maintains
the advantages observed in SpMV also for CG, achieving lower execution times,
reduced dynamic energy consumption, and more stable power profiles compared to
Ginkgo and AmgX across all tested matrices.

\section{Concluding Remarks and Future Work}
\label{concl}

This study demonstrates that optimizing numerical kernels and communication strategies in multi-GPU environments not only improves runtime performance, but also significantly reduces the energy footprint of large-scale sparse linear solvers. The results confirm that \texttt{BootCMatchGX} is competitive with, and in most cases superior to, established libraries such as Ginkgo and NVIDIA AmgX. Specifically, it achieves higher energy efficiency in SpMV and CG computations, while delivering a clear advantage in PCG thanks to its preconditioner design.

From an application perspective, these findings highlight the importance of considering energy efficiency as a key metric in the development of next-generation numerical libraries. Sustainable high-performance computing will increasingly depend on software that balances performance scalability with optimized energy use.

Future research will be directed toward:
\begin{itemize}
\item Developing efficient and scalable AMG preconditioners that leverage mixed-precision arithmetic, with the goal of further reducing both execution time and energy consumption while preserving numerical robustness.
\item  Validating the library’s energy benefits in multidisciplinary application domains, such as large-scale PDE simulations and scientific machine learning.
\end{itemize}

\section*{Acknowledgement}

The authors would like to thank the developers of GPowerU for providing an effective tool for measuring GPU energy consumption, developed within the framework of the European TEXTAROSSA project and employed as the basis for this work. We also gratefully acknowledge Marcel Koch, from the Scientific Computing Center (SCC) at Karlsruhe Institute of Technology and developer of Ginkgo, for his support in its use.

\bibliographystyle{ieeetr} 
\bibliography{biblioenergy}

\end{document}